\documentclass{article}
\usepackage[utf8]{inputenc}
\usepackage{authblk}
\usepackage{setspace}
\usepackage[margin=1.25in]{geometry}
\usepackage{graphicx}
\graphicspath{ {./figures/} }
\usepackage{subcaption}
\usepackage{amsmath}
\usepackage{lineno}
\usepackage{hyperref}
\usepackage{color}

\usepackage[style=nejm, 
citestyle=numeric-comp,
sorting=none]{biblatex}
\title{Dew harvesting grass:\\
{\Large{role of epicuticular wax in regulating condensation dynamics}}} 

\author[1]{Bashra Mahamed}
\author[2]{Francis James Dent}
\author[3]{Robert Simpson}
\author[4]{Nicola Weston}
\author[5]{Fanny Nascimento Costa}
\author[2*]{Sepideh Khodaparast}

\affil[1]{School of Physics and Astronomy, University of Leeds, Leeds LS2 9JT, UK.}
\affil[2]{School of Mechanical Engineering, University of Leeds, Leeds LS2 9JT, UK.}
\affil[3]{School of Chemical and Process Engineering, University of Leeds, Leeds LS2 9JT, UK.}
\affil[4]{Nanoscale \& Microscale Research Centre (nmRC), University of Nottingham, University Park, Nottingham NG7 2RD, UK.}
\affil[5]{The Bragg Centre for Materials Research, Sir William Henry Bragg Building, University of Leeds, Leeds LS2 9JT, UK.}

\affil[*]{Address correspondence to: s.khodaparast@leeds.ac.uk}

\date{}

\begin{document}

\maketitle

\begin{abstract}

Identification and characterization of natural dew collecting models is instrumental for the inspiration, design and development of engineered dew harvesting systems. 
Short low growing grass is one of the most ubiquitous and proficient examples of natural dew harvesting, owing to its large surface area, small thermal capacity, structured rough surface and proximity to ground level.
Here, we provide a closer look at the formation, growth, and dynamics of microscale dew droplets on the surface of wheatgrass leaves, investigating the role of epicuticular wax.
The wheatgrass leaf exhibits biphilic properties emerging from the hydrophilic lamina covered by hydrophobic wax microsculptures. 
As a result, the regulation of the dew formation and condensation dynamics is largely governed by the arrangement and density of epicuticular wax micromorphologies.
At moderate subcooling levels (4-10~$^{\circ}$C below the dew point), we observe drop-wise condensation on the superhydrophobic adaxial side, while significant flooding and film condensation usually appear on abaxial surfaces with lower wax coverage.
On the adaxial side of the leaves, the fairly uniform coverage of the hydrophobic epicuticular wax crystals on the hydrophilic background promotes drop-wise condensation nucleation while facilitating droplet mobility.
Frequent coalescence of multiple droplets of 5~-~20~\textmu m diameter results in self-propelled departure events, creating free potential sites for new nucleation.
The findings of this study advance our understanding of dew formation on natural surfaces while providing inspiration and guidance for the development of sustainable functional microstructured coatings for various drop-wise condensation applications. 
\end{abstract}


\section{Introduction}

Water scarcity is an emerging global challenge exacerbated by anthropogenic changes that disrupt weather patterns and strain freshwater resources.
By 2025, around half of the world's population is estimated to live in areas facing water shortage problems, leading to a displacement of around 700 million people by 2030.\cite{WaterAid:2018}
Insufficient supply met with increasing demand hence makes freshwater availability one of the United Nations' top priorities for sustainable development.\cite{Unicef,UNESCO:2009} 
The discovery and development of new technologies that tap into alternative water sources is thus needed to complement existing surface and ground water supplies such as rivers, reservoirs, lakes, and underground aquifers.
With the atmosphere of the Earth holding around 10$\%$ of the world's freshwater, atmospheric water harvesting technology (AWH) is an emerging strategy for decentralized and sustainable water production.\cite{Liu:2017,Zhou:2020} 
Across different geographical regions, strategies to collect water from the atmosphere in terms of dew, fog, and drizzle have been estimated to produce around 40~$\%$ extra fresh water.\cite{Beysens:2007,Beysens:2024,Liu:2020,Liu:2020a,Muselli:2009,Ritter:2019,Tomaszkiewicz:2015,Meunier:2016}
Fog harvesting is effective in coastal and mountainous fog-prone regions where liquid droplets of water are trapped and harvested from the atmosphere.\cite{Kennedy:2024}
Alternatively, in arid and semi-arid areas, dew harvesting provides a water extraction process for irrigation and fresh water consumption by taking advantage of radiative cooling below the dew point.\cite{DaSilva:2021,Beysens:2024}

As a viable form of AWH, the effectiveness of dew harvesting technologies is constrained by environmental and technological factors. 
As dew formation relies on condensation nucleation and manipulation of small water droplets on the surface,\cite{Beysens:1995} the physical topography and interfacial energy of the surface across scales play crucial roles in the efficiency of the designs.\cite{Khodaparast:2021,Seo:2016,Gerasopoulos:2018}
Heterogeneous condensation may occur in film- or drop-wise regimes depending on the thermodynamic supersaturation level and properties of the condenser surface, with drop-wise being favoured in AWH technologies because of its higher collection efficiency and lower water loss due to evaporation.\cite{ROSE:1981}
Recent developments have introduced a range of innovative micro/nanostructuring techniques to promote drop-wise condensation on the surface and increase the overall efficiency of the approach.\cite{Boreyko:2009,Miljkovic:2012,Chen:2011,Tang:2021}
Across diverse applications, efficient designs are required to facilitate cyclic nucleation, growth and droplet shedding, which maintain a quasi-steady operation for drop-wise condensation cooling devices and dew collectors.\cite{Hengyi:2022,Seo:2016}
While successful designs across nano- to macroscales have been proposed and trialed, advancing the overall performance, scalability, sustainability and affordability of dew harvesting materials remains an active field of research.\cite{Wang:2023}

Examples of successful biological systems have been the main source of inspiration in the last two decades to achieve a rational design for the approach, structure and materials of AWH technologies.\cite{Andrews:2011,Roth:2012,Ju:2012,Barthlott:2016,Dai:2018,Bhushan:2019,Gao:2024,Wang:2023,Kennedy:2024} 
In nature, dew harvesting relies on a passive approach, where moisture from the air condenses on surfaces that are chilled during the night through radiative cooling;\cite{Beysens:2007} these passive systems do not require external energy inputs and are largely dependent on environmental conditions.\cite{Monteith:1965,Ritter:2019} 
Interfacial condensation occurs as the surface temperature drops below the dew point, allowing water collection on surfaces such as plants, rocks, or engineered materials with high thermal emissivity.\cite{Zhuang:2017,Li:2002,Liu:2020a,Beysens:2007}
Being a major critical abiotic factor that modulates plant growth, health and productivity, water has been a primary evolutionary force in the development of plants, resulting in outstanding functionalities to facilitate alternative water access and inhibit moisture loss.\cite{Mauseth:2006,Norgaard:2010,Roth:2012,Lusa:2014,Barthlott:2016,Rockwell:2022,Bobe:2006,Prochazka:2024}
Passive dew harvesting is thus especially efficient across diverse species of plants inhabiting arid environments,\cite{Malik:2014,Yu:2022} with identified examples in variety of dessert plants providing water sources for arthropods,\cite{Broza:1979,Roth:2012} hierarchical three-dimensional arrangements in \textit{Cotula fallax} plant\cite{Andrews:2011} and dew harvesting cacti species.\cite{Malik:2015}

As a dominant group of organisms on the earth, plants occupy a large surface area on our planet; 
their leaves act as natural water regulators, collecting and controlling moisture under changing climatic circumstances.\cite{Hakeem:2023,Dawson:2018}
Barthlott \emph{et al}. estimated that the surface area covered by water repellent plant leaves, such as those found in grasses, is equal to about 50$\%$ of the total surface of the earth.\cite{Barthlott:2017}
The external surface of the majority of plant leaves comprises a cuticle that is covered by a thin waxy layer, acting as a hydrophobic barrier.\cite{Eglinton:1967,Shepherd:2006}
This epicuticular layer is composed of crystalline waxes, typically mixtures of aliphatic hydrocarbons and their derivatives, such as primary and secondary alcohols, ketones, fatty acids, and aldehydes.\cite{Koch:2008}
The significance of the micromorphology of epicuticular wax on the wetting behaviour of the plant leaves has been extensively studied in relation to micro/macroscopic contact with water droplets,\cite{Holloway:1969,Ensikat:2011,Szczepanski:2017,Bhushan:2010,Chakraborty:2019,Muhammad:2020,Dent:2024} however, the link between the form and distribution of the epicuticular wax and the collection of microscopic dew droplets on the leaf is not yet fully understood.
Studies on dew formation on plants are largely focused on macroscale field measurements and modeling of foliar water uptake and their ecological consequences at various environmental conditions.\cite{Hirst:1954,Sharma:1976,Hughes:1994,Gerlein:2018,Muselli:2009,Jacobs:2008,Agam:2006,Liu:2020a,Xiao:2009}

\begin{figure}[ht!]
    \centering
    \includegraphics[width=0.65\linewidth]{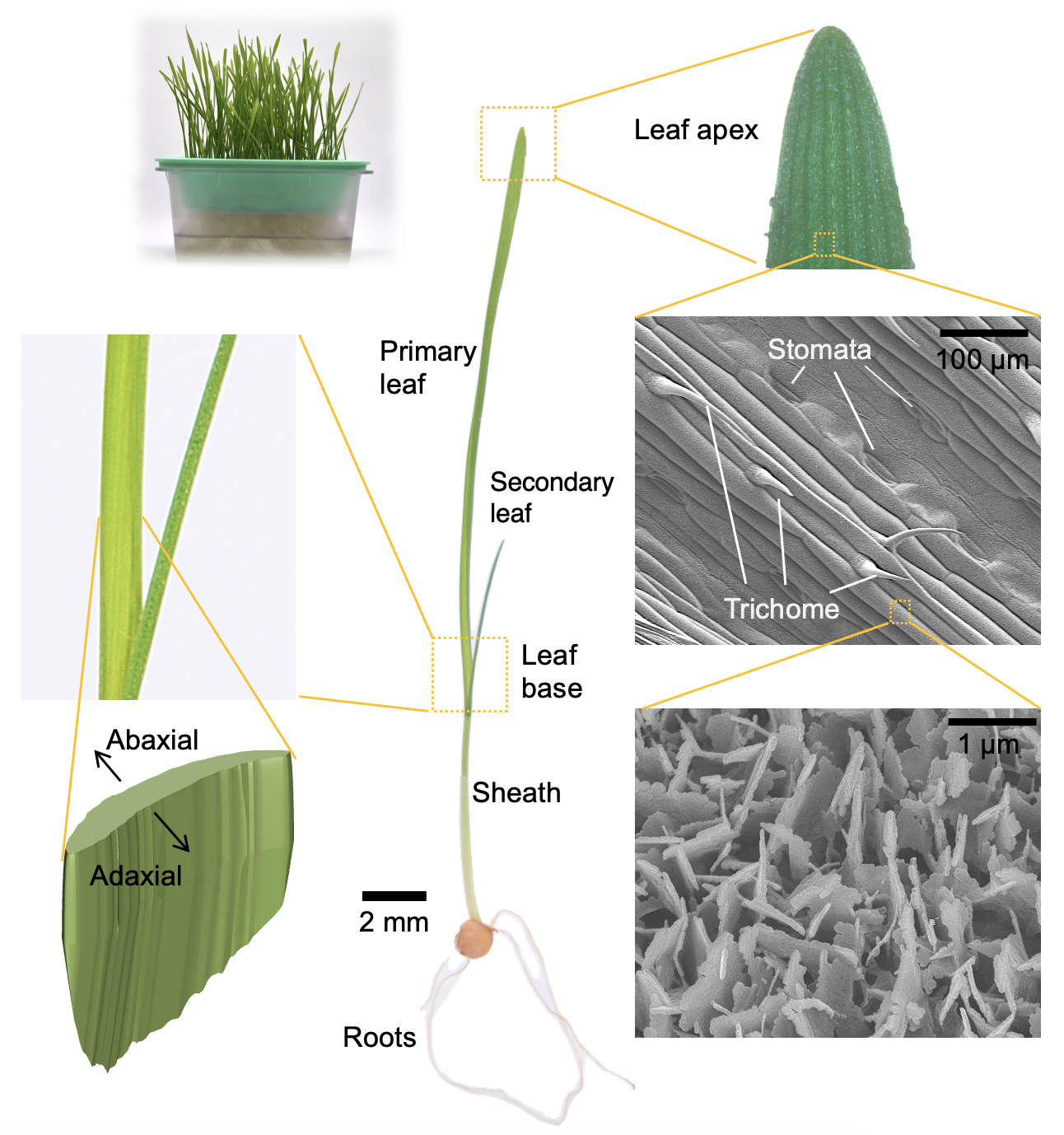}
    \caption{Key features of wheatgrass leaf used in this study. 
    Images highlight key features of the wheatgrass plant, adaxial and abaxial sides, leaf regions and multiscale surface architecture.}
    \label{fig:leaf}
\end{figure}

Previous works have highlighted the role of plant architecture or millimetric features on the surface of plants, such as ridges, hairs, spines and trichomes, in collecting fog and dew in different species of plants.\cite{Kennedy:2024,Roth:2012,Ju:2012,Waseem:2021,Bin:2022}
Quantitative analysis of larger millimetric dew and fog droplets has also been reported on the grass \emph{Holcus lanatus}\cite{Hughes:1994} and \emph{Stipagrostis sabulicola}.\cite{Roth:2012} 
In comparison, the significance of the interfacial plant wax microstructure during dew formation especially at early stages of nucleation and growth is less studied.\cite{Nath:2019,Xu:2022}
Here, we investigate several key aspects of the topographic design of natural leaves and their impact on the nucleation and growth of microscopic dew droplets at different environmental conditions. 
Wheatgrass leaves are chosen as a model low-growing grass with distinct wax coverage across different regions of the leaves because of their high yield, simplicity and ease of growth for laboratory study.
Special attention is paid to the impact of epicuticular wax micromorpholgy and coverage on the dynamics of drop-wise condensation through \textit{in situ} microscopy.\cite{Zheng:2021}
Wheatgrass is a common example of efficient natural dew harvesting material due to its surface structure and larger interfacial area relative to its ground surface footprint (Fig.~\ref{fig:leaf}).\cite{Long:1955} 
We employ active cooling to allow in situ visualization of condensation dynamics and correlate findings with the surface topography. 
We observe the formation of microscale dew droplets in different regions at the tip, centre and base along the leaves, distinguishing the abaxial and adaxial sides (Fig.~\ref{fig:leaf}).
Natural micro- and nanoarchitectures are readily reproducible on a variety of hard, soft, and porous material,\cite{Jeffree:1975,Jetter:1994,Koch:2006,Dent:2024} thus providing opportunities for the design and fabrication of a new generation of scalable dew harvesting coatings.
Furthermore, the dynamics of leaf surface interaction with environmental water resources discussed here is closely linked to the adsorption of liquid pesticides on plants,\cite{Yu:2009,Damak:2016} making this research important for both natural and agricultural ecosystems.\cite{Dawson:2018,Liu:2020a,Cavallaro:2022}

\section{Results and discussions}

The formation of dew on grass leaves in their natural setting occurs due to effective passive radiative cooling in the darkness resulting in a surface temperature decrease to below that of the dew point.\cite{Beysens:2024}
To perform a quantitative systematic analysis of dew dynamics under well-controlled thermodynamic conditions, we performed measurements through active cooling of wheatgrass leaves by a Peltier device.\cite{Dent:2022}
This approach allows us to investigate the role of surface topography in isolation from variant environmental conditions.
We report results of condensation experiments at subcooling levels in the range of $\Delta T_{\text{c}} = 2-10~^{\circ}$C, comparable to observations in natural environments and reports of recent emerging engineering technologies.\cite{Nath:2019,Haechler:2021,Malik:2014,Beysens:2024}


\subsection{Surface topography}
\begin{figure}[ht!]
    \centering
    \includegraphics[width=0.88\linewidth]{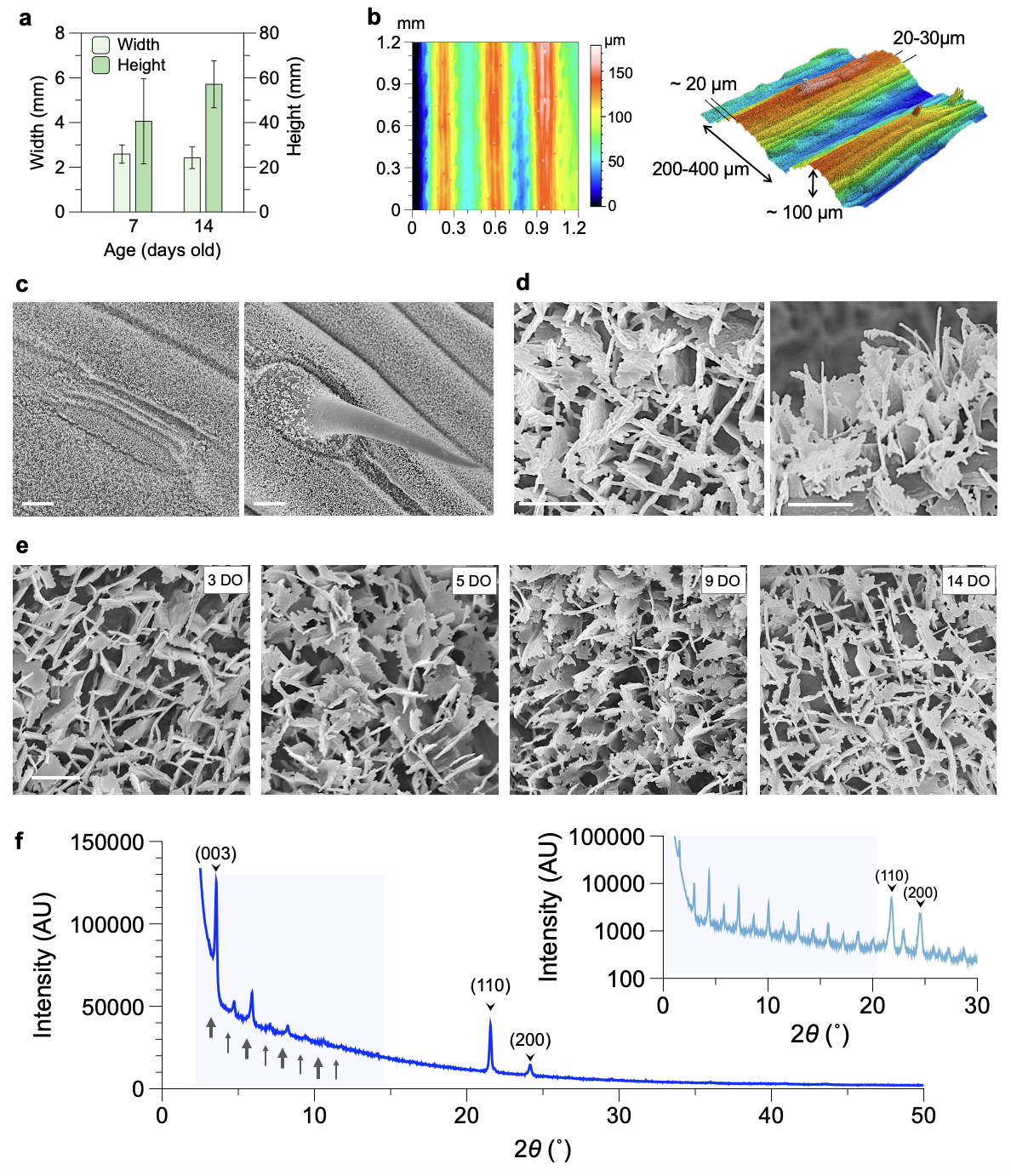}
    \caption{Multi-scale topography of wheatgrass leaf.
    (a) The average height and width, measured for leaves up to 7 and 14 days of age. 
    Error bars represent the standard deviation of the population.
    (b) 2D contour plots and 3D LSCM images show distribution of elevation across an example 1.2 x 1.2~mm$^{2}$ region of the leaf.
    (c) The adaxial surface of the leaf, including the stomata (left), is densely covered by the wax crystal plates, except for the trichomes (right).
    (d) SEM close-up images of the wax crystals show their standing platelets structures with ragged sides resembling a two-dimensional fir tree. 
    (e) SEM images on the adaxial apex regions of leaves at different ages.
    Scale bars correspond to 10~\textmu m in {\bf{c}} and 1~\textmu m in {\bf{d}} and {\bf{e}}.
    (f) X-ray diffraction intensity for the wax collected from the leaf through dissolution in chloroform showing the orthorhombic crystal structure.
    Inset shows the diffraction intensity measured for pure 1-Octacosanol powder.
    Small-angle regions are highlighted in blue.}
    \label{fig:topography}
\end{figure}

The surface of wheatgrass leaf embodies a combination of multi-scale directional and isotropic topographies, ranging from a few millimeters down to nanometers (Figs.~\ref{fig:leaf}, 2). 
The fresh leaves of about 2~mm width grow predominantly in length, reaching an average height of 40 to 60~mm from 7 to 14 days, respectively (Fig.~\ref{fig:topography}a).  
Longitudinal veins and cell arrangements create compound surface undulations with dimensions of 10-100's~\textmu m, see Fig.~\ref{fig:topography}b and Fig.~S1.\cite{Vofely:2019,Kumar:2019}
At a significantly smaller scale, interfacial nano/micromorphologies appear as a result of molecular self-assemblies and microscale aggregation of epicuticular waxes, central to the focus of this work. 

The adaxial surface of the leaves, including the stomata, is covered by densely packed wax crystals (Fig.~\ref{fig:topography}c, left);
the only exception is the trichome surface (Fig.~~\ref{fig:topography}c, right) which exhibits decreasing wax crystal density from the base to tip. 
The wax micromorphology appears as randomly oriented platelets of approximately 50 nm thickness with ragged edges similar to those observed on leaf blades of \emph{T. aestivum} and \emph{Odosicyos} (Fig.~\ref{fig:topography}d).\cite{Barthlott:1998,Koch:2006}
These platelets are 0.5-1 \textmu m long and 0.5-2 \textmu m high, and generally appear to be taller and more densely packed compared to those found on mature wheat (\emph{T. aestivum}) blades.\cite{Koch:2006} 
The wax crystals were observed to be well developed shortly after the emergence of the primary leaf.
Starting from 5 days of leaf growth, randomly oriented wax platelets formed a populated hydrophobic mesh on the leaf, see Fig.~\ref{fig:topography}e.
Most of the analysis presented here was therefore performed on freshly harvested 7-day-old blades with consistent wax coverage.
No significant variation in wax coverage and crystal morphology was found in the apex region of the leaves of different heights harvested at a specific age, see Fig.~S2.

X-ray diffraction analyses of the extracted wheatgrass leaf wax demonstrated the orthorhombic structure of the wax crystal (Fig.~\ref{fig:topography}f), manifested by the two intense head-group spacing peaks at 4.1 $\AA$ and 3.7 $\AA$, see Fig.~S3.
Epicuticular wax platelets extracted from wheat blades were previously found to be mainly composed of primary alcohols, in particular 1-octacosanol, which accounts for up to 70$\%$ of the overall mass composition.\cite{Koch:2006,Bianchi:1977}
XRD measurement of 1-octacosanol is provided in Fig.~\ref{fig:topography}f for comparison.
The highlighted long-spacing region ($2\theta \leq 20^{\circ}$) of the graph demonstrates equally spaced peaks at alternating high and low intensity, which were attributed to the alkyl-alkyl and hydroxyl-hydroxyl boundaries in the bilayer structure of long-chain fatty alcohols.\cite{Koch:2006}
Further analysis of the long-spacing peaks confirmed a relatively wider bilayer d-spacing in the extracted wheatgrass wax crystals compared to the pure 1-octacosanol, possibly due to the existence of longer chain aliphatic compounds in the extracted wax, see Fig.~S2. 



\subsection{Surface wetting}

\begin{figure}[ht!]
    \centering
    \includegraphics[width=\linewidth]{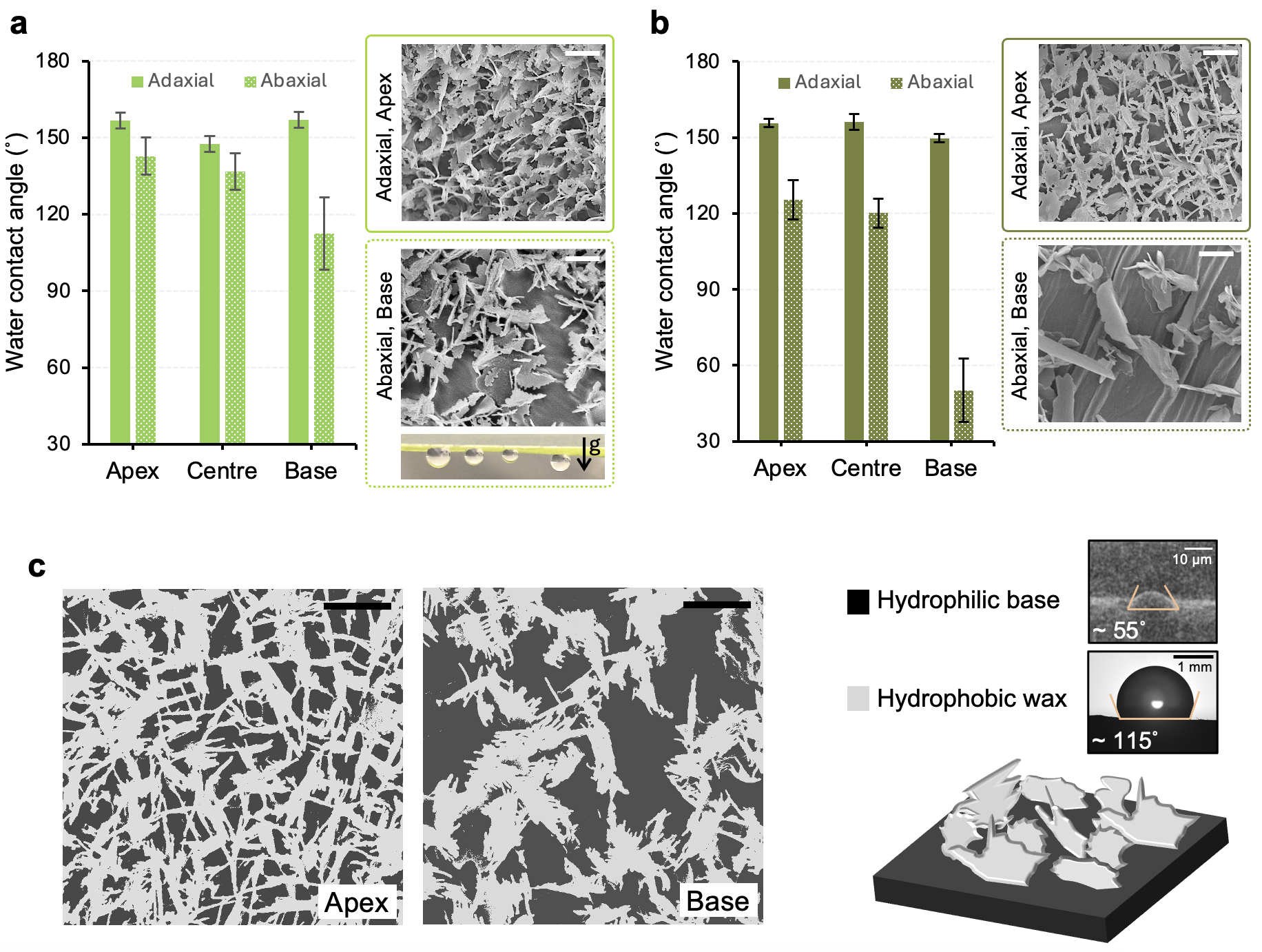}
    \caption{Impact of epicuticular wax coverage on the wetting behaviour of the wheatgrass leaf.
    Water contact angle and representative SEM images across different regions on the adaxial and abaxial sides of (a) freshly harvested 7 days old and (b) dehydrated 14 days old leaves. 
    Water contact angles are presented on the apex, centre and base of the wheatgrass leaves. 
    The surface of the abaxial side is typically sticky towards water droplets.  
    Error bar represent the standard deviation of average measurements on 5 independent leaves.
    Scale bars correspond to 1 \textmu m in all images in {\bf{a}} and {\bf{b}}.
    (c) Binary SEM images of the apex and base regions on the abaxial side of the 7 days old fresh leaves.
    Hydrophilic base and hydrophobic waxes are demonstrated in dark and light colors.
    Average contact angles are measured using ESEM and optical micro-imaging on the hydrophilic areas between the wax crystals and a flat 1-octacosanol coating, respectively.}
    \label{fig:CA}
\end{figure}

Noticeably different wetting behavior was observed across different regions of the leaf, captured via static water contact angle measurements. 
The comparison between the measured water contact angle for the adaxial and abaxial sides is presented in Fig.~\ref{fig:CA}a and Fig.~\ref{fig:CA}b for freshly harvested and dehydrated leaves, respectively.
The microsculptures on the adaxial side of the leaves yield slippery superhydrophobic properties with static water contact angles $\geq 150^{\circ}$, similar to observations reported for other plant species.\cite{Holloway:1969,Koch:2006}
In contrast, the wetting behaviour varies from the apex to the base region on the abaxial side of the leaves; the region at the base of the leaves is less hydrophobic showing a decrease in contact angle values in both fresh leaves (Fig.~\ref{fig:CA}a) and dehydrated leaves (Fig.~\ref{fig:CA}b).
This behavior was found to be directly correlated with the epicuticular wax surface coverage, noticeable in the example SEM images presented in Fig.~\ref{fig:CA}.
In addition to the reduced hydrophobicity, the base regions demonstrated a sticky wetting behaviour with highly pinned droplets as indicated in the inset image of Fig.~\ref{fig:CA}a.


The reduced hydrophobicity in the base region was more significant in the aged dehydrated leaves with more aggregated and dispersed wax crystal platelets, leaving extended wax-free regions on the surface and resulting in significantly smaller static contact angles $\leq 90^{\circ}$.
While long-term exposure to drought in natural environments may cause changes in wax composition and often an increase in the cuticular wax content to minimize water loss in plants,\cite{Zhu:2013,Bi:2017} the reduction in the epicuticular wax coverage in the lab-grown aged dehydrated leaves may be due to environmental mechanical abrasion or other aggregation mechanisms triggered by water loss and cell shrinkage in the leaf cuticles.\cite{Hoad:1992,Koch:2006}
The considerable non-homogeneity in the epicuticular wax crystal coverage resulted in significant pinning of the air-water contact line in the wax-free spaces, reported on other natural and laboratory-made coatings of heterogeneous chemical or physical texture.\cite{Joanny:1990,Bauer:2012,Bartolo:2006}
A slight reduction in epicuticular wax crystal coverage was also observed in all regions of the adaxial side in aged dehydrated leaves, however, the consequences on the overall macroscopic wetting behaviour of the surface was not significant (Fig.~\ref{fig:CA}b). 

The enhanced surface wettability in regions with low wax coverage suggests a biphilic composition as a result of the hydrophilic leaf surface decorated with superficial hydrophobic wax platelets, as illustrated in the top-view SEM images and the schematic in Fig.~\ref{fig:CA}c.\cite{Mishchenko:2013}
ESEM and optical microscopy images in Fig.~\ref{fig:CA}c show contact angle measurements on flat coatings made up of 1-octacosanol ($\approx 115 \pm 2^{\circ}$) and on the bare leaf lamina ($\approx 55 \pm 5^{\circ}$) corroborating the biphilic property of the wheatgrass leaf surface. 
As nucleation and growth of condensation droplets on surfaces with heterogeneous wetting properties are influenced by the dimensions and arrangement of wetting and non-wetting features, we investigated dew dynamics in leaf regions with distinctive wax coverage guided by our wetting analysis.\cite{Kim:2024} 

\subsection{Dew dynamics}
Dew formation on the wheatgrass leaf is an example of heterogeneous condensation on a rough biphilic surface.
Hydrophilic substrates exhibit lower energy barriers to condensation nucleation.
Continuous nucleating sites spread, however, resulting in the film-wise condensation regime on such surfaces that is undesirable to both cooling and water harvesting applications.\cite{ROSE:1981}
Hydrophobic interfacial microstructures are essential for the thermodynamic phase change to result in the formation of isolated quasi-spherical droplets, referred to as drop-wise condensation or breath figure formation.\cite{Beysens:1995}
While the overall transition between the film- to drop-wise condensation can often be simply predicted based on the macro wetting properties of the surface and the thermodynamic supersaturation level, dynamics of condensation growth is governed by interfacial interactions at micro- and nanoscale.
We performed a series of condensation experiments on different regions of leaves with varying wax coverage at discrete saturation levels established by horizontally mounting leaf sections on a Peltier device at a set subcooling level $\Delta T_{\text{c}}$.
The wax-free elongated trichomes (Fig.~\ref{fig:topography}c) are expected to play a significant role in defining the fate of dew droplets only at later stages of condensation (Fig.~S4), therefore, the present study of earlier stages nucleation and growth primarily focuses on the micropatterned leaf lamina.\cite{Ju:2012,Waseem:2021}
The following results highlight the role of epicuticular wax micromorphology in prescribing the outcome of condensation on the leaves and establishing complex dynamics that is often engineered in deliberately designed nanostructured surfaces.

\subsubsection{Condensation regimes}

Tracking the average grayscale intensity in top-view microscopy images after subtracting the background (initial reference image) offers a facile approach to visualize the ensemble effects of the nucleation and growth of dew droplets as well as different regimes of growth and coalescence.
\textit{In situ} bright-field microscopy was performed in reflection mode, thereby readily capturing the formation of curved microdroplet interfaces was readily captured as increased intensity in the images. 

\begin{figure}
    \centering
    \includegraphics[width=0.8\linewidth]{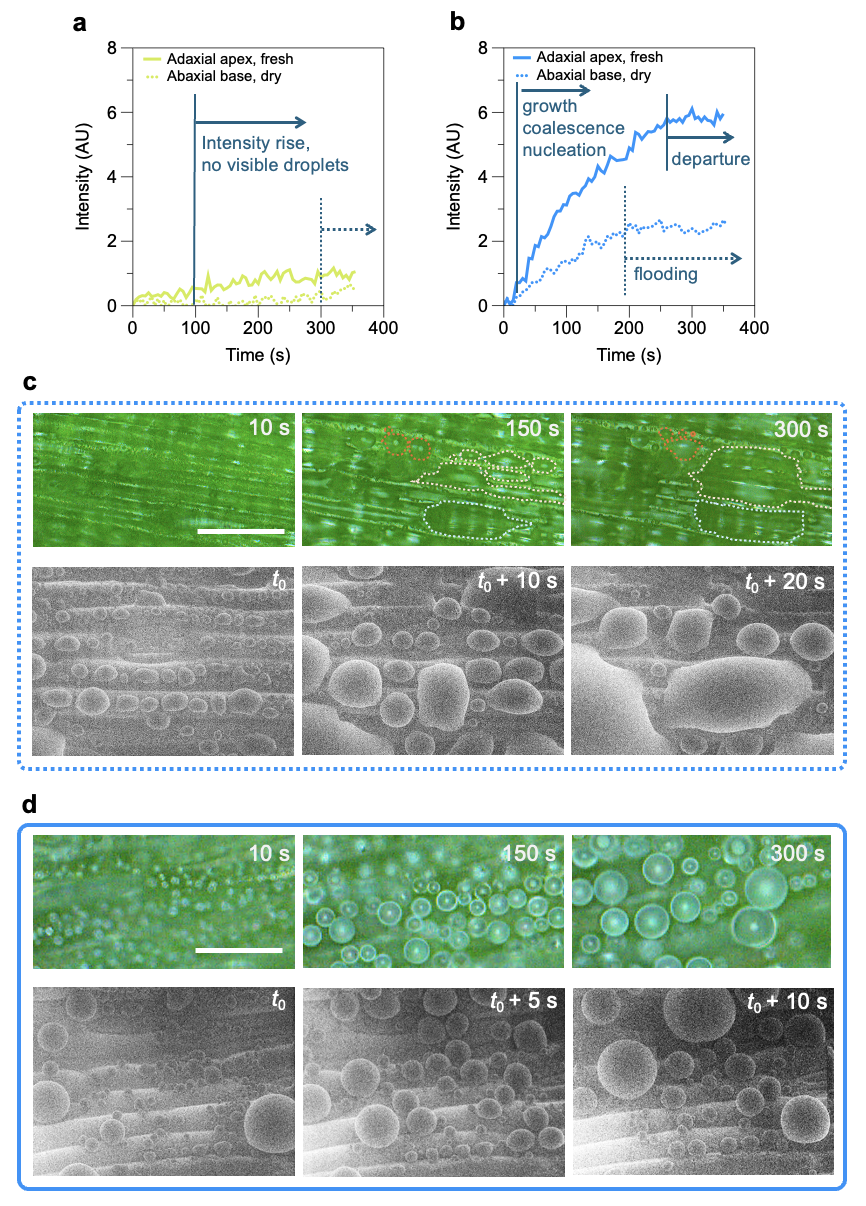}
    \caption{Condensation regimes on regions of varying hydrophobicity on wheatgrass leaves.
    Graphs show distinctive relative grayscale intensity evolutions at constant subcooling at (a) $\Delta T_{\text c} = 2$~$^{\circ}$C and (b) $\Delta T_{\text c} = 10$~$^{\circ}$C on the adaxial apex and abaxial base regions of fresh and dehydrated leaves.   
    (a) No optically resolvable drop-wise condensation is found at small subcooling within the duration of the experiments.
    The onset of gradual increase in the intensity was observed at earlier times on the apex region of fresh leaves possibly due to moisture adsorption or the formation of sub-micron droplets on the surface. 
    (b) At larger subcooling, drop-wise condensation is observed on both surfaces.
    Optical microscopy (top) and ESEM (bottom) images of drop-wise condensation evolution on the (c) abaxial base and (d) adaxial apex of the wheatgrass leaves.
    Scale bar refers to 50~\textmu m in optical and ESEM images in {\bf{c}} ad {\bf{d}}.}
    \label{fig:ADAB}
\end{figure}

The example results of nucleation and growth of condensation droplets are presented in Fig.~\ref{fig:ADAB} for the most disparate variations of the wax coverage regions between fresh and dehydrated leaves. 
The apex region of the adaxial side (fresh) is significantly more hydrophobic than the base of the abaxial side in dry leaves due to the higher intensity of wax coverage, with water contact angles measured at approximately 155~$^{\circ}$C compared to 45~$^{\circ}$C from Fig.~\ref{fig:CA}. 
The graphs in Fig.~\ref{fig:ADAB}a and Fig.~\ref{fig:ADAB}b compare the results of nucleation at smaller ($\Delta T_{\text{c}} = 2$~$^{\circ}$C) and larger ($\Delta T_{\text{c}} = 10$~$^{\circ}$C) subcooling for these samples.
For $\Delta T_{\text{c}} = 2$~$^{\circ}$C, no clear air-water interface or droplet formation was observed regardless of the overall surface wetting characteristic (Fig.~\ref{fig:ADAB}a).
Environmental SEM analysis at small subcooling showed rare appearance of unstable small condensation droplets which immediately disappeared. 
In contrast, the two regions on the abaxial and adaxial sides of the leaves promoted different regimes of condensation at larger subcooling.
Initial drop-wise condensation on the base regions spread on the more hydrophilic surface, generating larger flooded regions (Fig.~\ref{fig:ADAB}c, top) manifested by a longer region of constant intensity in Fig.~\ref{fig:ADAB}b, see Vid-S1.
Relatively larger number of spherical micro droplets were found on the superhydrophobic adaxial side of the leaf tip, with regular coalescence upon growth, see Vid-S2.
These droplets maintained their spherical shape from the early stages of condensation throughout the growth regimes (Fig.~S5).
Droplet coalescence reduced the overall number of droplets, leading to a slower increase of captured reflected light intensity at intermediate times.
This resulted in a quasi-constant intensity regime, where coalescence results in the departure of a significant number of droplets from the surface, which were subsequently replaced by a new generation of small droplets.\cite{Boreyko:2009}

In summary, we observed three distinctive condensation dynamic regimes across the different regions of wheatgrass leaves, captured through the ensemble relative grayscale analysis:
(i) moisture adsorption with no apparent air-water interface formation at smaller subcooling ($\Delta T_{\text c} = 2$~$^{\circ}$C) across all analyzed samples, 
(ii) drop-wise condensation leading to non-spherical growth, interface pinning and local flooding in less hydrophobic abaxial base regions, and 
(iii) nucleation and spherical growth of dew droplets in superhydrophobic regions. 
Increasing the subcooling level induced larger initial nucleation densities and droplet growth rates, resulting in significant droplet coalescence and departure in (iii). 



\subsubsection{Kinetics of drop-wise condensation}

\begin{figure}
    \centering
    \includegraphics[width=0.9\linewidth]{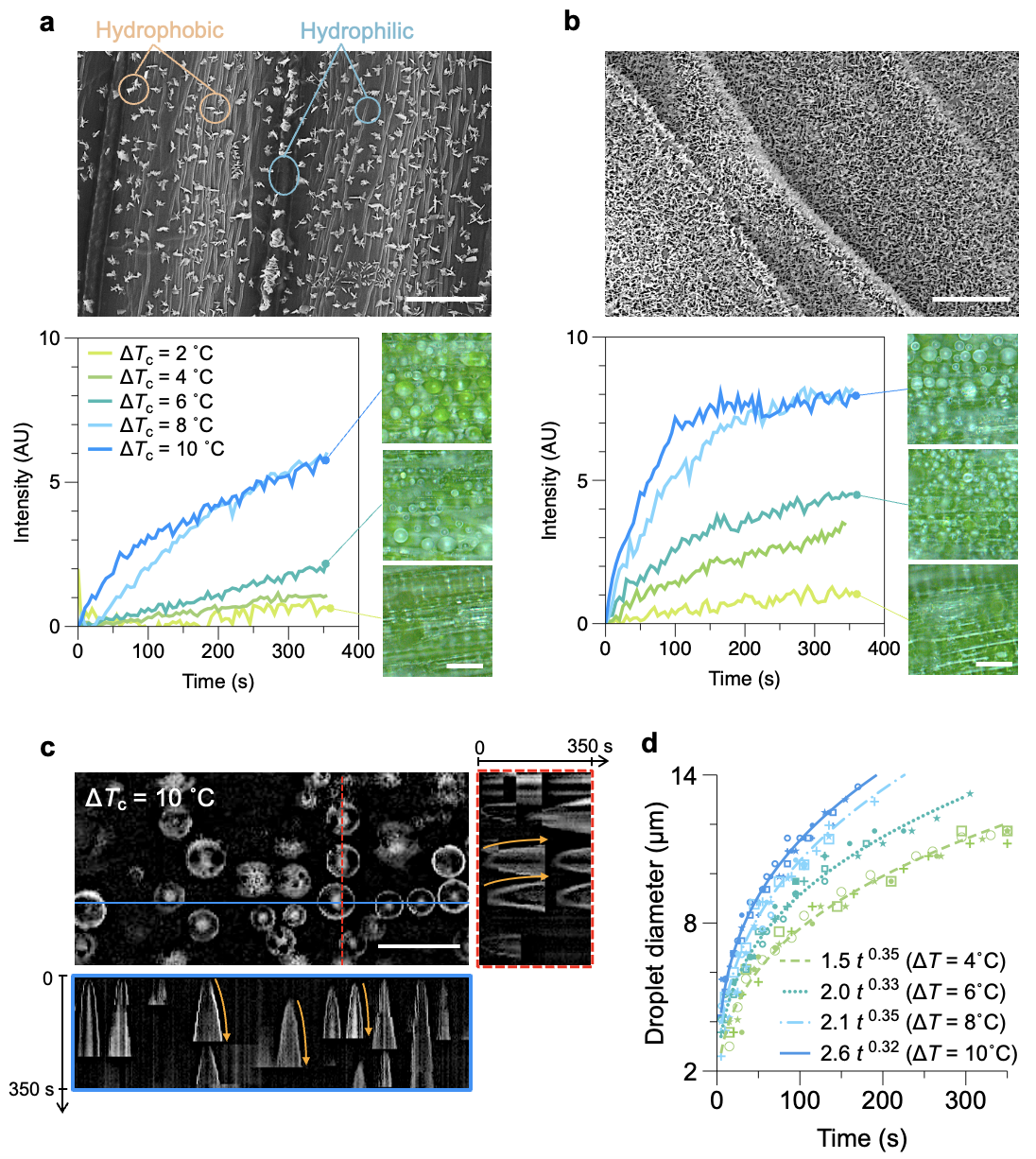}
    \caption{Impact of wax crystal coverage on the nucleation and growth of condensation droplets on the abaxial surface of fresh 7 days old leaves. 
    SEM images show sparse wax crystals on the base (a) relative to the higher wax coverage on the apex region (b).
    Graphs show the relative mean grayscale intensity evolution during subcooling experiments at varied subcooling level denoted by $\Delta T_{\text c}$ for the base and the apex regions.
    Optical microscopy images refer to the final state in the cooling experiment at 350 s. 
    (c) Orthogonal time-strip analysis demonstrates the symmetric growth of condensation droplets on the apex region by tracking the relative grayscale intensity along horizontal and vertical lines in a cropped view of microscopy images. 
    Arrows show the direction of growth along the detected bright interface of the droplets.
    (d) Growth of individual droplets on the apex region ({\bf{b}}, {\bf{c}}) is governed by molecular diffusion before coalescence occurs, captured by a power-law relationship between the droplet and time. 
   Scale bars refer to 10 \textmu m for SEM images in {\bf{a}} and {\bf{b}}, and 50 \textmu m in optical micrographs presented in {\bf{a, b}} and {\bf{c}}.}
    \label{fig:AB-comparison}
\end{figure}

{\bf{Nucleation and growth.}} While the adaxial surface of the fresh leaf remains superhydrophobic throughout with approximately constant wax coverage, the epicuticular wax distribution on the abaxial region varies across the blade length showing consistent increase of surface coverage from the base to the tip.
Figs.~\ref{fig:AB-comparison}a,b highlight the wax coverage variation and subsequent effects on the condensation regime between the base and apex abaxial regions of fresh 7 day old leaves, respectively. 

Drop-wise nucleation and growth of dew droplets was observed on both regions for $\Delta T_{\text c} \geq 4$~$^{\circ}$C throughout the analysis, however the number of detected microdroplets was significantly smaller on the base region dispersed wax crystal aggregates (Fig.~\ref{fig:AB-comparison}a) compared to the uniformly coated apex region (Fig.~\ref{fig:AB-comparison}b).
While the hydrophilic base promoted nucleation, denser packing of hydrophobic wax crystals was essential for the growth of a larger number of spherical droplets, resulting in overall larger intensities detected on the apex region.
On the abaxial base region of fresh leaves droplets grew into larger non-spherical shapes, which generally yielded lower detected light reflection intensity in the microscopy images.
See Fig.~\ref{fig:AB-comparison}a and b and refer to Vid-S3 for a comparative view of nucleation an growth on the two sites on the abaxial side.
Additionally, the uniform wax coverage on the tip further enhanced droplet coalescence and mobility that is reflected in the plateau region in intensity-time graph at later times for higher levels of subcooling (Fig.~\ref{fig:AB-comparison}b). 

Spherical droplets were formed from the early stages of microscopic detection (diameter $\geq$ 4~\textmu m) on the adaxial apex region of the leaves (Fig.~S5). 
These droplets were observed to grow symmetrically in time, as captured by an example time-strip analysis in Fig.~\ref{fig:AB-comparison}c where the time evolution of relative grayscale intensities is tracked along orthogonal lines.
The growth of drops across the two orthogonal axes leads to the appearance of horizontal and vertical cones that expand over time. 
The interruption in the growth of the cone base diameter is due to the coalescence of neighboring droplets and the dislocation or departure of the newly formed droplet. 
Although accurate quantitative tracking of the average droplet growth within the field of the view was not possible due to the out-of-focus effects resulted from the microscopic leaf surface undulation, measurement of single droplet diameters in time follow the well-known power law, $D \propto t^{1/3}$ predicted for growth of condensation droplets dominated by molecular diffusion (Fig.~\ref{fig:AB-comparison}d).\cite{Viovy:1988,Fritter:1991}
Longer periods of diffusion-dominant growth was captured for the lowest subcooling level ($\Delta T_{\text c} = 4$~$^{\circ}$ C) as a consequence of the lower initial nucleation density and coalescence, and slower droplet growth. 

\begin{figure}
    \centering
    \includegraphics[width=0.9\linewidth]{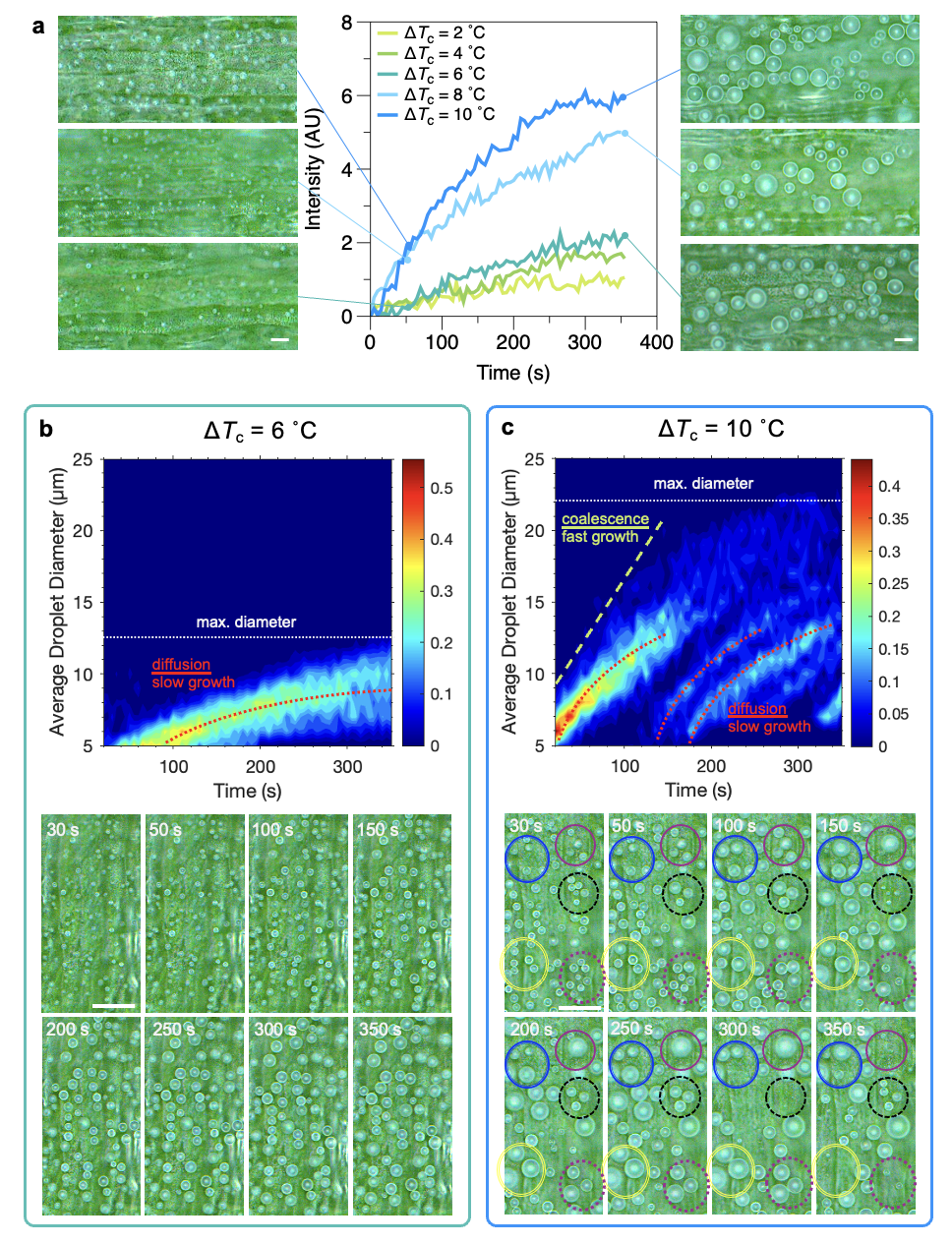}
    \caption{Dew formation on the adaxial apex region of the 7 days old wheatgrass leaves. 
    (a) Evolution of grayscale intensity in time at different levels of subcooling.
    Micrographs correspond to cropped regions of the original images at 50~s and 350~s. 
    (b) Population density of average droplet diameter at $\Delta T_{\text{c}}$ = 4~$^{\circ}$C highlighting the diffusion dominated growth regimes.
    (c) Growth, coalescence, departure and recondensation events during a subcooling experiment at $\Delta T_{\text{c}}$ = 10~$^{\circ}$C.
    Droplet coalescence and departure enables re-condensation appearing as new populations of droplets growing through molecular diffusion between the coalescence events.
    Circles highlight regions of the image where droplet departure and re-condensation occur.
    Scale bars correspond to 20 \textmu m in all microscopy images.}
    \label{fig:condensation-SH}
\end{figure}

{\bf{Coalescence and departure.}} On the adaxial side of the leaf with unifrom wax coverage, drop-wise condensation was observed for $\Delta T_{\text c} \geq 4$~$^{\circ}$C.
On these superhydrophobic regions of the leaves, decreasing the subcooling temperature was found to impact the initial nucleation density, growth and thus the probability and outcome of the droplet coalescence events (Fig.~\ref{fig:condensation-SH}). 
Analysis of population density of droplet diameter was performed over cropped in-focus regions of the images to quantitatively describe aspects of the condensation dynamics (Fig.~S6).

At subcooling $\Delta T_{\text c} = 4-6$~$^{\circ}$C, the grayscale variation increases at a much slower rate compared to larger subcooling, $\Delta T_{\text c} = 8-10$~$^{\circ}$C (Fig.~\ref{fig:condensation-SH}a).
Inset images in Fig.~\ref{fig:condensation-SH}b for $\Delta T_{\text c} = 6$~$^{\circ}$C show isolated droplets growing almost exclusively via diffusion of water vapour from the air.
Little to no coalescence between neighbouring droplets occur within the experimental time frame.
The time evolution of droplet diameter density in Fig.~\ref{fig:condensation-SH}b demonstrates this with a concentrated narrow range in the average diameter over time mainly governed by the well-described diffusion mechanism.\cite{Viovy:1988}
Slower droplet growth and limited coalescence at moderately small subcooling restricts the average and maximum droplet diameters at about 9 and 12 \textmu m, respectively.

At higher subcooling level ($\Delta T_{\text c} = 8-10$~$^{\circ}$C), both number density and growth rates of condensation droplets increased significantly, leading to frequent coalescence of the droplets captured by the plateau region in the grayscale intensity graph after 200 s from the start of the experiments, demonstrating that a balance between the old interfaces disappearing and newly generated interfaces is achieved.
Fig.~S7 compares examples of droplet dynamics at smaller and larger subcooling through time-strip analysis. 
The plot of diameter population density versus time shows a much faster initial droplet growth over the first 150 s, followed by multiple new families of droplets being nucleated, compare Figs.~\ref{fig:condensation-SH}b with ~\ref{fig:condensation-SH}c.
The coalescence of the droplets was captured as an increase in the slope and a widening of the intensity-time curves.\cite{Fritter:1991,Dent:2022}
A close-up view of the microscopic images of the condensation further links quantitative light intensity measurements to the kinetics of nucleation and growth of condensation droplets. 
Microscale droplets were observed to nucleate across the microsctructured surface and readily coalesce upon contact; examples of coalescence events are highlighted in Fig.~\ref{fig:condensation-SH}c. 
Between the 100~s and 150~s time frames, coalescent events between multiple droplets and high mobility led to the energetic departure from the surface, freeing up space for the nucleation of new droplet families; see Vid-S2 and Vid-S4.
Due to the larger initial nucleation density and high mobility of the interface on the nanostructured adaxial surface of the leaf, most of the coalescence events involved three or more droplets with diameters well below 10~\textmu m, comparable to some of the most efficient engineered surface designs for continuous drop-wise condensation and shedding.\cite{Boreyko:2009,Chen:2011,Anand:2012,Miljkovic:2012,Miljkovic:2013,Enright:2014a,Jonathan:2024,Sarkiris:2024}
This effective dynamic nucleation, growth, coalescence, departure and re-nucleation cycle keeps the maximum detected diameter below around 20~\textmu m within the experimental time frame.

\section{Conclusions}

Low-growing grasses are short in height, inhabiting areas close to the ground with limited wind.
Their large surface area, low thermal capacity, and microstructured surface enables them to be highly effective dew collectors in this environment.
By investigating the role of hydrophobic epicuticular wax crystals on condensation, we identified regimes of moisture adsorption and dew formation on the surface of wheatgrass leaves as a representative model.
The epicuticular wax, made up of orthorhombic crystal structures, generated interfacial hydrophobic micromorphologies in the form of extended platelets of about 10's nm thickness and 1's \textmu m length/height on the hydrophilic foundation of the leaf, resulting in biphilic wetting properties. 
On the abaxial side, the wax coverage wax reduced from the apex region towards the base of the leaf, resulting in reduced hydrophobicity and local transitions from drop-wise to film-wise condensation.
Regions of high wax coverage on the adaxial side of the leaves show slippery superhydrophobic properties that promotes effective drop-wise condensation through continuous cycles of nucleation, growth, coalescence and departure, and re-nucleation. 

The interfacial hydrophobic wax platelets appears as a promising choice of surface treatment for the generation and maintenance of effective drop-wise condensation and dew formation for the scalable fabrication of bio-inspired functional coatings, benefiting from availability and sustainability of plant waxes and the simplicity of available manufacturing techniques.\cite{Koch:2006a,Miljkovic:2012,Bayer:2020,Dent:2024}
The microscale surface features in the form of indented grooves or protruded trichomes are expected to have significant effect on the later stage dynamics of dew formation and collection of the larger droplets on the actual grass blades, alongside the leaf deformability and edge effects.
Future work on vertical standing grass blades will be essential to identify the fate of the dew droplets reported here on horizontal leaves.

\section{Materials and methods}

\subsection{Wheatgrass leaf growth}
Seeds of wheatgrass - first leaves of the common wheat plant (\emph{Triticum aestivum}) - were purchased from Pretty Wild Seeds, UK.
To optimize germination conditions, seeds were rinsed and allowed to soak for 3 days. 
The seeds were rinsed twice daily during the soaking period and then transferred to standard hydroponic sprouting trays before being placed in a growth chamber on day 4. 
The growth chamber was maintained at a relative humidity (RH) of 60\% and a temperature of 20°C ± 3°C with a scheduled 16/8 hour light/dark cycle to simulate natural daylight conditions and promote growth.
The sprouting trays consisted of a mesh top where the germinated seeds were spread and a bottom reservoir tray in which roots grew;
the water in the reservoir (root container) was changed twice daily in the morning and evening. 
Plant specimens were harvested at discrete growth stages shortly after the leaf emergence stage and up to 14 days after the initial germination conditions were started.
This growth period is significantly shorter than that of wheat plant crops grown for human food production,\cite{Koch:2006,Bianchi:1977} facilitating high throughput of fresh samples for analysis.
Specimens were sampled immediately prior to analysis to ensure the leaf blade remained in a hydrated state.
Analysis was carried out in the apex, center and base regions of the primary leaf shoot, on both adaxial and abaxial sides (Fig.~\ref{fig:leaf}).
Physical measurements of the grass leaves were taken at several stages of growth. 
The width and length of individual leaves were measured using a caliper gauge and a ruler. 
For dehydrated samples, wheatgrass leaves were removed from the growth chamber and left in a low humidity environment at RH $\approx$ 40$\%$ from day 10 to 14.

\subsection{Surface analysis}

{\bf{Leaf surface topography.}} Surface topography measurements were performed on fresh leaves using a Laser Scanning Confocal Microscope (LSCM) (Carl Zeiss LSM800) equipped with a 405 nm laser, with data exported to MountainsMap\textregistered software (Digital Surf).
Scanning electron microscopy (SEM) analysis was performed on freshly cut sections of the leaves mounted on a stub using double sided carbon tape. 
The samples were sputter coated with 10 nm platinum prior to imaging on a Hitachi SU8230 microscope operated at 2 kV accelerating voltage.

{\bf{Wax crystal structure.}} Crystal structures were analysed by X-ray diffraction (XRD), using a Bruker D8 diffractometer with monochromatic Cu K$\alpha$ radiation (2$\theta$ range 2.5$^{\circ}$–50$^{\circ}$, step size = 0.013$^{\circ}$).
Measurements were carried out over 12 hours to improve the signal-to-noise ratio. 
The epicuticular wax was dissolved in chloroform by submerging the leaves in the solvent for 10 seconds. 
The solution was then filtered and drop-casted on zero-background silicon substrates. 
Powder X-ray diffraction was also performed on 1-Octacosanol ($\geq$ 99$\%$) sourced from Sigma-Aldrich.

{\bf{Wetting analysis.}} Fresh wheatgrass leaves were harvested before every measurement and mounted on microscope glass slides using double sided tape.
Care was taken to ensure that the region of interest was not touched when securing the sample flat to the substrate.
Static contact angle measurements were performed using a customized micro-imaging setup and analyzed using the drop analysis plugin on ImageJ.\cite{Stalder:2006} 
The average value of the left and right static contact angles was measured for water droplets of 2~\textmu m volume, with five independent leaves tested for each condition. 
On each surface, the average contact angle for three droplets was calculated to account for the surface nonhomogeneity. 
Droplet deposition by pipetting was not possible on the superhydrophobic sections of leaves; therefore, water droplets were injected using a syringe and held attached to blunt needles above the surface. 
The measurements were then performed by gradually raising the level of the leaf surface to contact the water droplets.
The error bars on all presented graphs correspond to the standard deviations.

\subsection{\textit{In situ} condensation visualization}

{{\bf Optical microscopy.}} Condensation experiments were performed under monitored laboratory environmental conditions of temperature ($T_0$) and relative humidity (RH).
Fresh wheatgrass leaves were mounted on standard borosilicate glass coverslips of 24 mm × 24 mm with a thickness of 0.15 mm to ensure minimal thermal resistance.
The mounted leaf was placed on the Peltier device held at a constant temperature $T_\text{P}$ during each experiment, accurate to within ±0.5~°C.
Subcooling temperature was calculated based on the difference between the measured dew point and the set Peltier temperature, $\Delta T_{\text{c}} = T_{\text{dp}} - T_{\text{P}}$.
Condensation dynamics were monitored using an Olympus BX53M optical microscope (OM) in reflective mode, equipped with a long working distance objective (Olympus LMPLFLN 20X) and a digital CMOS camera (Basler ace acA2040-90uc). 
All kinetics were recorded and analyzed starting from the initial time the grass leaf was exposed to the desired subcooling, capturing images every 5 seconds with a nominal spatial resolution of 0.55~\textmu m per pixel.
Three independent runs were performed for each location, using newly harvested leaves to ensure reproducibility. 
The graphs in the article represent examples of various single runs at different locations and environmental conditions to improve visibility. 
Data analysis was performed in ImageJ on grayscale images by subtracting the initial image from the sequence and tracking the average grayscale intensity over time.
The nucleation and growth of condensation droplets resulted in the appearance of higher grayscale intensity in the background-subtracted images because of the high reflection at the water-air interface.
Although larger microscale surface topographies at the location of the veins produced out-of-focus effects across the images, quantifying the ensemble intensity of the reflected light allowed for effective tracking of the condensation kinetics from the nucleation and growth of new interfaces across the entire field of view.

In-focus image sections of sample experiments that displayed drop-wise condensation were analyzed by calculating the average droplet diameter.
Images were digitally enhanced to increase the contrast of droplet borders before running previously developed MATLAB (Mathworks, R2024b) code to automatically identify droplet perimeters across the time series.\cite{Dent:2023} 
The population density distribution of measured average droplet size in time was visualized using 3D mesh plots and the corresponding contour plots.
In each time frame, 1~\textmu m bins were created with normalized counts to create a density distribution.

{{\bf Environmental scanning electron microscopy.}} Early stage nucleation and growth of water condensation droplets were imaged using FEI Quanta 650 with a tungsten source.
Fresh samples were harvested prior to each analysis and mounted to angled conductive stubs placed on a Peltier module.
No conductive coating was used to view the condensation kinetics on the virgin grass surface.
The microscope chamber was purged and pumped down to 200~Pa before decreasing the Peltier temperature to 2~$^\circ$C, ensuring the relative humidity remained well below the saturation pressure.
The chamber pressure was slowly increased to increase the relative humidity with images taken at different times using a 5~kV accelerating voltage.

\section{Acknowledgment}
B.H. was supported by Leeds Institute of Fluid Dynamics (LIFD) and Yorkshire Consortium for Equity in Doctoral Education (YCEDE) through their summer internship programme.
F.J.D. is funded through an EPSRC Doctoral Prize Fellowship (EP/W524372/1).
Cost of the experimental work was partially supported by the Leeds Institute of Fluid Dynamics (LIFD).
The ESEM analysis was supported by the Engineering and Physical Sciences Research Council (EPSRC) under grant EP/L022494/1 and the University of Nottingham.

\newpage
\printbibliography

@article{Dent:2024,
	author = {Dent, Francis J. and Tyagi, Gunjan and Esat, Faye and Cabral, Jo{\~a}o T. and Khodaparast, Sepideh},
	doi = {https://doi.org/10.1002/adfm.202307977},
	eprint = {https://onlinelibrary.wiley.com/doi/pdf/10.1002/adfm.202307977},
	journal = {Advanced Functional Materials},
	number = {1},
	pages = {2307977},
	title = {Tuneable Topography and Hydrophobicity Mode in Biomimetic Plant-Based Wax Coatings},
	url = {https://onlinelibrary.wiley.com/doi/abs/10.1002/adfm.202307977},
	volume = {34},
	year = {2024}}

@article{Haechler:2021,
	author = {Iwan Haechler and Hyunchul Park and Gabriel Schnoering and Tobias Gulich and Mathieu Rohner and Abinash Tripathy and Athanasios Milionis and Thomas M. Schutzius and Dimos Poulikakos},
	doi = {10.1126/sciadv.abf3978},
	eprint = {https://www.science.org/doi/pdf/10.1126/sciadv.abf3978},
	journal = {Science Advances},
	number = {26},
	pages = {eabf3978},
	title = {Exploiting radiative cooling for uninterrupted 24-hour water harvesting from the atmosphere},
	url = {https://www.science.org/doi/abs/10.1126/sciadv.abf3978},
	volume = {7},
	year = {2021}}

@techreport{Unicef,
	author = {Department of Economic and Social Affairs, United Nations},
	institution = {Statistics Division},
	title = {Clean Water and Saniration},
	year = {2022}}

@article{Khodaparast:2021,
	author = {Khodaparast, Sepideh and Marcos, Julius and Sharratt, William N. and Tyagi, Gunjan and Cabral, Jo{\~a}o T.},
	date = {2021/01/12},
	doi = {10.1021/acs.langmuir.0c02821},
	isbn = {0743-7463},
	journal = {Langmuir},
	journal1 = {Langmuir},
	journal2 = {Langmuir},
	month = {01},
	number = {1},
	pages = {230--239},
	publisher = {American Chemical Society},
	title = {Surface-Induced Crystallization of Sodium Dodecyl Sulfate (SDS) Micellar Solutions in Confinement},
	type = {doi: 10.1021/acs.langmuir.0c02821},
	url = {https://doi.org/10.1021/acs.langmuir.0c02821},
	volume = {37},
	year = {2021},
	year1 = {2021}}

@article{Mishchenko:2013,
	author = {Mishchenko, Lidiya and Khan, Mughees and Aizenberg, Joanna and Hatton, Benjamin D.},
	doi = {https://doi.org/10.1002/adfm.201300418},
	eprint = {https://advanced.onlinelibrary.wiley.com/doi/pdf/10.1002/adfm.201300418},
	journal = {Advanced Functional Materials},
	number = {36},
	pages = {4577-4584},
	title = {Spatial Control of Condensation and Freezing on Superhydrophobic Surfaces with Hydrophilic Patches},
	url = {https://advanced.onlinelibrary.wiley.com/doi/abs/10.1002/adfm.201300418},
	volume = {23},
	year = {2013}}

@article{Kim:2024,
	author = {Kim, Moon-Kyung and Sett, Soumyadip and Hoque, Muhammad Jahidul and Kim, Euichel and Ahn, Junyoung and Miljkovic, Nenad},
	date = {2024/08/20},
	doi = {10.1021/acs.langmuir.4c02247},
	isbn = {0743-7463},
	journal = {Langmuir},
	journal1 = {Langmuir},
	journal2 = {Langmuir},
	month = {08},
	number = {33},
	pages = {17767--17778},
	publisher = {American Chemical Society},
	title = {Fundamental Limits of the Spatial Control of Heterogeneous Nucleation on Biphilic Surfaces},
	type = {doi: 10.1021/acs.langmuir.4c02247},
	url = {https://doi.org/10.1021/acs.langmuir.4c02247},
	volume = {40},
	year = {2024},
	year1 = {2024}}

@article{Roth:2012,
	author = {Roth-Nebelsick, A. and Ebner, M. and Miranda, T. and Gottschalk, V. and Voigt, D. and Gorb, S. and Stegmaier, T. and Sarsour, J. and Linke, M. and Konrad, W.},
	doi = {10.1098/rsif.2011.0847},
	eprint = {https://royalsocietypublishing.org/doi/pdf/10.1098/rsif.2011.0847},
	journal = {Journal of The Royal Society Interface},
	number = {73},
	pages = {1965-1974},
	title = {Leaf surface structures enable the endemic Namib desert grass Stipagrostis sabulicola to irrigate itself with fog water},
	url = {https://royalsocietypublishing.org/doi/abs/10.1098/rsif.2011.0847},
	volume = {9},
	year = {2012}}

@article{Kennedy:2024,
	author = {Kennedy, Brook S. and Boreyko, Jonathan B.},
	doi = {https://doi.org/10.1002/adfm.202306162},
	eprint = {https://advanced.onlinelibrary.wiley.com/doi/pdf/10.1002/adfm.202306162},
	journal = {Advanced Functional Materials},
	number = {35},
	pages = {2306162},
	title = {Bio-Inspired Fog Harvesting Meshes: A Review},
	url = {https://advanced.onlinelibrary.wiley.com/doi/abs/10.1002/adfm.202306162},
	volume = {34},
	year = {2024}}

@article{ROSE:1981,
	author = {J.W. Rose},
	doi = {https://doi.org/10.1016/0017-9310(81)90026-0},
	issn = {0017-9310},
	journal = {International Journal of Heat and Mass Transfer},
	number = {2},
	pages = {191-194},
	title = {Dropwise condensation theory},
	url = {https://www.sciencedirect.com/science/article/pii/0017931081900260},
	volume = {24},
	year = {1981}}

@article{Malik:2015,
	author = {F T Malik and R M Clement and D T Gethin and D Beysens and R E Cohen and W Krawszik and A R Parker},
	doi = {10.1088/1748-3190/10/3/036005},
	journal = {Bioinspiration \& Biomimetics},
	month = {apr},
	number = {3},
	pages = {036005},
	publisher = {IOP Publishing},
	title = {Dew harvesting efficiency of four species of cacti},
	url = {https://dx.doi.org/10.1088/1748-3190/10/3/036005},
	volume = {10},
	year = {2015}}

@article{Norgaard:2010,
	author = {N{\o}rgaard, Thomas and Dacke, Marie},
	date = {2010/07/16},
	doi = {10.1186/1742-9994-7-23},
	id = {N{\o}rgaard2010},
	isbn = {1742-9994},
	journal = {Front. Zool.},
	number = {1},
	pages = {23},
	title = {Fog-basking behaviour and water collection efficiency in Namib Desert Darkling beetles},
	url = {https://doi.org/10.1186/1742-9994-7-23},
	volume = {7},
	year = {2010}}

@article{Wang:2023,
	author = {Wang, Yi and Zhao, Weinan and Han, Mei and Xu, Jiaxin and Tam, Kam Chiu},
	date = {2023/07/01},
	doi = {10.1038/s44221-023-00109-1},
	id = {Wang2023},
	isbn = {2731-6084},
	journal = {Nature Water},
	number = {7},
	pages = {587--601},
	title = {Biomimetic surface engineering for sustainable water harvesting systems},
	url = {https://doi.org/10.1038/s44221-023-00109-1},
	volume = {1},
	year = {2023}}

@article{Koch:2006a,
	author = {Koch, Kerstin and Barthlott, Wilhelm},
	date = {2006/11/01},
	doi = {10.1177/1934578X0600101123},
	isbn = {1934-578X},
	journal = {Natural Product Communications},
	journal1 = {Natural Product Communications},
	month = {2024/10/25},
	number = {11},
	pages = {1934578X0600101123},
	publisher = {SAGE Publications Inc},
	title = {Plant Epicuticular Waxes: Chemistry, Form, Self-Assembly and Function},
	type = {doi: 10.1177/1934578X0600101123},
	url = {https://doi.org/10.1177/1934578X0600101123},
	volume = {1},
	year = {2006},
	year1 = {2006}}

@article{Viovy:1988,
	author = {Viovy, Jean Louis and Beysens, Daniel and Knobler, Charles M.},
	doi = {10.1103/PhysRevA.37.4965},
	issue = {12},
	journal = {Phys. Rev. A},
	month = {Jun},
	numpages = {0},
	pages = {4965--4970},
	publisher = {American Physical Society},
	title = {Scaling description for the growth of condensation patterns on surfaces},
	url = {https://link.aps.org/doi/10.1103/PhysRevA.37.4965},
	volume = {37},
	year = {1988}}

@article{Enright:2014a,
	author = {Enright, Ryan and Miljkovic, Nenad and Sprittles, James and Nolan, Kevin and Mitchell, Robert and Wang, Evelyn N.},
	date = {2014/10/28},
	doi = {10.1021/nn503643m},
	isbn = {1936-0851},
	journal = {ACS Nano},
	journal1 = {ACS Nano},
	journal2 = {ACS Nano},
	month = {10},
	number = {10},
	pages = {10352--10362},
	publisher = {American Chemical Society},
	title = {How Coalescing Droplets Jump},
	type = {doi: 10.1021/nn503643m},
	url = {https://doi.org/10.1021/nn503643m},
	volume = {8},
	year = {2014},
	year1 = {2014}}

@article{Fritter:1991,
	author = {Fritter, Daniela and Knobler, Charles M. and Beysens, Daniel A.},
	doi = {10.1103/PhysRevA.43.2858},
	issue = {6},
	journal = {Phys. Rev. A},
	month = {Mar},
	numpages = {0},
	pages = {2858--2869},
	publisher = {American Physical Society},
	title = {Experiments and simulation of the growth of droplets on a surface (breath figures)},
	url = {https://link.aps.org/doi/10.1103/PhysRevA.43.2858},
	volume = {43},
	year = {1991}}

@article{Dent:2022,
	author = {Dent, Francis J. and Harbottle, David and Warren, Nicholas J. and Khodaparast, Sepideh},
	date = {2022/06/15},
	doi = {10.1021/acsami.2c05635},
	isbn = {1944-8244},
	journal = {ACS Applied Materials \& Interfaces},
	journal1 = {ACS Applied Materials \& Interfaces},
	journal2 = {ACS Appl. Mater. Interfaces},
	month = {06},
	number = {23},
	pages = {27435--27443},
	publisher = {American Chemical Society},
	title = {Temporally Arrested Breath Figure},
	type = {doi: 10.1021/acsami.2c05635},
	url = {https://doi.org/10.1021/acsami.2c05635},
	volume = {14},
	year = {2022},
	year1 = {2022}}

@article{Dent:2023,
	author = {Dent, Francis J. and Harbottle, David and Warren, Nicholas J. and Khodaparast, Sepideh},
	doi = {10.1039/D2SM01650H},
	issue = {15},
	journal = {Soft Matter},
	pages = {2737-2744},
	publisher = {The Royal Society of Chemistry},
	title = {Exploiting breath figure reversibility for in situ pattern modulation and hierarchical design},
	url = {http://dx.doi.org/10.1039/D2SM01650H},
	volume = {19},
	year = {2023}}

@article{Bianchi:1977,
	author = {Giorgio Bianchi and Maria Corbellini},
	doi = {https://doi.org/10.1016/S0031-9422(00)86700-X},
	issn = {0031-9422},
	journal = {Phytochemistry},
	number = {7},
	pages = {943-945},
	title = {Epicuticular wax of Triticum aestivum demar 4},
	url = {https://www.sciencedirect.com/science/article/pii/S003194220086700X},
	volume = {16},
	year = {1977}}

@article{Jeffree:1975,
	author = {Jeffree, C. E. and Baker, E. A. and Holloway, P. J.},
	journal = {New Phytol.},
	pages = {539--549},
	title = {Ultrastruture and recrystallization of plant epicuticular waxes},
	volume = {75},
	year = {1975}}

@article{Jetter:1994,
	author = {Jetter, J. and Riederer, M.},
	journal = {Planta},
	pages = {257--270},
	title = {Epicuticular crystals of nonacosan-10-ol: In-vitro reconstitution and factors influencing crystal habits},
	volume = {195},
	year = {1994}}

@article{Xiao:2009,
	author = {H. Xiao and R. Meissner and J. Seeger and H. Rupp and H. Borg},
	doi = {https://doi.org/10.1016/j.jhydrol.2009.08.006},
	issn = {0022-1694},
	journal = {Journal of Hydrology},
	number = {1},
	pages = {43-49},
	title = {Effect of vegetation type and growth stage on dewfall, determined with high precision weighing lysimeters at a site in northern Germany},
	url = {https://www.sciencedirect.com/science/article/pii/S0022169409004806},
	volume = {377},
	year = {2009}}

@article{Tang:2021,
	author = {Tang, Yu and Yang, Xiaolong and Li, Yimin and Lu, Yao and Zhu, Di},
	date = {2021/11/24},
	doi = {10.1021/acs.nanolett.1c01584},
	isbn = {1530-6984},
	journal = {Nano Letters},
	journal1 = {Nano Letters},
	journal2 = {Nano Lett.},
	month = {11},
	number = {22},
	pages = {9824--9833},
	publisher = {American Chemical Society},
	title = {Robust Micro-Nanostructured Superhydrophobic Surfaces for Long-Term Dropwise Condensation},
	type = {doi: 10.1021/acs.nanolett.1c01584},
	url = {https://doi.org/10.1021/acs.nanolett.1c01584},
	volume = {21},
	year = {2021},
	year1 = {2021}}

@article{Hengyi:2022,
	doi = {https://doi.org/10.1002/adma.202110079},
	eprint = {https://onlinelibrary.wiley.com/doi/pdf/10.1002/adma.202110079},
	journal = {Advanced Materials},
	number = {12},
	pages = {2110079},
	title = {Materials Engineering for Atmospheric Water Harvesting: Progress and Perspectives},
	url = {https://onlinelibrary.wiley.com/doi/abs/10.1002/adma.202110079},
	volume = {34},
	year = {2022}}

@article{Seo:2016,
	author = {Seo, Donghyun and Lee, Junghun and Lee, Choongyeop and Nam, Youngsuk},
	date = {2016/04/11},
	doi = {10.1038/srep24276},
	id = {Seo2016},
	isbn = {2045-2322},
	journal = {Scientific Reports},
	number = {1},
	pages = {24276},
	title = {The effects of surface wettability on the fog and dew moisture harvesting performance on tubular surfaces},
	url = {https://doi.org/10.1038/srep24276},
	volume = {6},
	year = {2016}}

@article{Kumar:2019,
	author = {Kumar, Dhinesh and Kellogg, Elizabeth A.},
	doi = {https://doi.org/10.1111/nph.15491},
	eprint = {https://nph.onlinelibrary.wiley.com/doi/pdf/10.1111/nph.15491},
	journal = {New Phytologist},
	number = {3},
	pages = {1260-1267},
	title = {Getting closer: vein density in C4 leaves},
	url = {https://nph.onlinelibrary.wiley.com/doi/abs/10.1111/nph.15491},
	volume = {221},
	year = {2019}}

@article{Anand:2012,
	author = {Anand, Sushant and Paxson, Adam T. and Dhiman, Rajeev and Smith, J. David and Varanasi, Kripa K.},
	date = {2012/11/27},
	doi = {10.1021/nn303867y},
	isbn = {1936-0851},
	journal = {ACS Nano},
	journal1 = {ACS Nano},
	journal2 = {ACS Nano},
	month = {11},
	number = {11},
	pages = {10122--10129},
	publisher = {American Chemical Society},
	title = {Enhanced Condensation on Lubricant-Impregnated Nanotextured Surfaces},
	type = {doi: 10.1021/nn303867y},
	url = {https://doi.org/10.1021/nn303867y},
	volume = {6},
	year = {2012},
	year1 = {2012}}

@article{Chen:2011,
	doi = {https://doi.org/10.1002/adfm.201101302},
	eprint = {https://onlinelibrary.wiley.com/doi/pdf/10.1002/adfm.201101302},
	journal = {Advanced Functional Materials},
	number = {24},
	pages = {4617-4623},
	title = {Nanograssed Micropyramidal Architectures for Continuous Dropwise Condensation},
	url = {https://onlinelibrary.wiley.com/doi/abs/10.1002/adfm.201101302},
	volume = {21},
	year = {2011}}

@article{Sarkiris:2024,
	author = {Sarkiris, Panagiotis and Constantoudis, Vassilios and Ellinas, Kosmas and Lam, Cheuk Wing Edmond and Milionis, Athanasios and Anagnostopoulos, John and Poulikakos, Dimos and Gogolides, Evangelos},
	doi = {https://doi.org/10.1002/adfm.202306756},
	eprint = {https://onlinelibrary.wiley.com/doi/pdf/10.1002/adfm.202306756},
	journal = {Advanced Functional Materials},
	number = {1},
	pages = {2306756},
	title = {Topography Optimization for Sustainable Dropwise Condensation: The Critical Role of Correlation Length},
	url = {https://onlinelibrary.wiley.com/doi/abs/10.1002/adfm.202306756},
	volume = {34},
	year = {2024}}

@article{Boreyko:2009,
	author = {Boreyko, Jonathan B. and Chen, Chuan-Hua},
	doi = {10.1103/PhysRevLett.103.184501},
	issue = {18},
	journal = {Phys. Rev. Lett.},
	month = {Oct},
	numpages = {4},
	pages = {184501},
	publisher = {American Physical Society},
	title = {Self-Propelled Dropwise Condensate on Superhydrophobic Surfaces},
	url = {https://link.aps.org/doi/10.1103/PhysRevLett.103.184501},
	volume = {103},
	year = {2009}}

@article{Vofely:2019,
	auid = {ORCID: 0000-0003-0378-7112; ORCID: 0000-0002-0369-8606; ORCID: 0000-0002-4308-5580},
	author = {V{\H o}f{\'e}ly, R{\'o}za V and Gallagher, Joseph and Pisano, Grace D and Bartlett, Madelaine and Braybrook, Siobhan A},
	cin = {Nat Plants. 2018 Nov;4(11):855. doi: 10.1038/s41477-018-0310-y. PMID: 30390087},
	copyright = {{\copyright}2018 The Authors. New Phytologist {\copyright}2018 New Phytologist Trust.},
	crdt = {2018/10/04 06:00},
	date = {2019 Jan},
	dcom = {20200121},
	dep = {20181003},
	doi = {10.1111/nph.15461},
	edat = {2018/10/04 06:00},
	gr = {R21 AT003396/AT/NCCIH NIH HHS/United States},
	issn = {1469-8137 (Electronic); 0028-646X (Print); 0028-646X (Linking)},
	jid = {9882884},
	journal = {New Phytol},
	jt = {The New phytologist},
	language = {eng},
	lid = {10.1111/nph.15461 {$[$}doi{$]$}},
	lr = {20240922},
	mh = {Anisotropy; Cell Shape; Image Processing, Computer-Assisted; *Phylogeny; *Plant Cells; Plant Epidermis/cytology; Plant Leaves/*cytology; Plants/genetics},
	mhda = {2020/01/22 06:00},
	month = {Jan},
	number = {1},
	oto = {NOTNLM},
	own = {NLM},
	pages = {540--552},
	phst = {2018/07/03 00:00 {$[$}received{$]$}; 2018/08/21 00:00 {$[$}accepted{$]$}; 2018/10/04 06:00 {$[$}pubmed{$]$}; 2020/01/22 06:00 {$[$}medline{$]$}; 2018/10/04 06:00 {$[$}entrez{$]$}; 2019/06/20 00:00 {$[$}pmc-release{$]$}},
	pii = {NPH15461},
	pl = {England},
	pmc = {PMC6585845},
	pmcr = {2019/06/20},
	pmid = {30281798},
	pst = {ppublish},
	pt = {Journal Article; Research Support, Non-U.S. Gov't; Research Support, U.S. Gov't, Non-P.H.S.},
	sb = {IM},
	status = {MEDLINE},
	title = {Of puzzles and pavements: a quantitative exploration of leaf epidermal cell shape.},
	volume = {221},
	year = {2019}}

@article{Meunier:2016,
	author = {D. Meunier and D. Beysens},
	doi = {https://doi.org/10.1016/j.atmosres.2016.03.014},
	issn = {0169-8095},
	journal = {Atmospheric Research},
	pages = {65-72},
	title = {Dew, fog, drizzle and rain water in Baku (Azerbaijan)},
	url = {https://www.sciencedirect.com/science/article/pii/S0169809516300643},
	volume = {178-179},
	year = {2016}}

@article{DaSilva:2021,
	author = {Corraide da Silva, Larissa and Oliveira Filho, Delly and Acioli Imbuzeiro, Hewlley Maria and Barros Monteiro, Paulo Marcos},
	doi = {10.2166/ws.2021.242},
	eprint = {https://iwaponline.com/ws/article-pdf/22/1/697/991673/ws022010697.pdf},
	issn = {1606-9749},
	journal = {Water Supply},
	month = {08},
	number = {1},
	pages = {697-706},
	title = {Analysis of different condensing surfaces for dew harvesting},
	url = {https://doi.org/10.2166/ws.2021.242},
	volume = {22},
	year = {2021}}

@book{UNESCO:2009,
	author = {UNESCO World Water Assessment Programme},
	publisher = {UNESDOC Digital Library},
	title = {Water in a changing world: the United Nations world water development report 3},
	year = {2009}}

@techreport{WaterAid:2018,
	author = {WaterAid},
	institution = {WaterAid},
	title = {The Water Gap -- The State of the World's Water 2018},
	year = {2018}}

@article{Beysens:1995,
	author = {D. Beysens},
	doi = {https://doi.org/10.1016/0169-8095(95)00015-J},
	issn = {0169-8095},
	journal = {Atmospheric Research},
	number = {1},
	pages = {215-237},
	title = {The formation of dew},
	url = {https://www.sciencedirect.com/science/article/pii/016980959500015J},
	volume = {39},
	year = {1995}}

@article{Beysens:2007,
	author = {D. Beysens and O. Clus and M. Mileta and I. Milimouk and M. Muselli and V.S. Nikolayev},
	doi = {https://doi.org/10.1016/j.energy.2006.09.021},
	issn = {0360-5442},
	journal = {Energy},
	note = {Third Dubrovnik Conference on Sustainable Development of Energy, Water and Environment Systems},
	number = {6},
	pages = {1032-1037},
	title = {Collecting dew as a water source on small islands: the dew equipment for water project in Bis˘evo (Croatia)},
	url = {https://www.sciencedirect.com/science/article/pii/S0360544206002684},
	volume = {32},
	year = {2007}}

@article{Liu:2020,
	author = {Liu, Shuhao and Zheng, Jeremy and Hao, Li and Yegin, Yagmur and Bae, Michael and Ulugun, Beril and Taylor, Thomas Matthew and Scholar, Ethan A. and Cisneros-Zevallos, Luis and Oh, Jun Kyun and Akbulut, Mustafa},
	date = {2020/05/13},
	doi = {10.1021/acsami.9b18928},
	isbn = {1944-8244},
	journal = {ACS Applied Materials \& Interfaces},
	journal1 = {ACS Applied Materials \& Interfaces},
	journal2 = {ACS Appl. Mater. Interfaces},
	month = {05},
	number = {19},
	pages = {21311--21321},
	publisher = {American Chemical Society},
	title = {Dual-Functional, Superhydrophobic Coatings with Bacterial Anticontact and Antimicrobial Characteristics},
	type = {doi: 10.1021/acsami.9b18928},
	url = {https://doi.org/10.1021/acsami.9b18928},
	volume = {12},
	year = {2020},
	year1 = {2020}}

@article{Hakeem:2023,
	author = {Hakeem, Sadia and Ali, Zulfiqar and Saddique, Muhammad Abu Bakar and Merrium, Sabah and Arslan, Muhammad and Habib-ur-Rahman, Muhammad},
	date = {2023/02/27},
	doi = {10.1186/s12870-023-04123-z},
	id = {Hakeem2023},
	isbn = {1471-2229},
	journal = {BMC Plant Biology},
	number = {1},
	pages = {115},
	title = {Leaf wettability and leaf angle affect air-moisture deposition in wheat for self-irrigation},
	url = {https://doi.org/10.1186/s12870-023-04123-z},
	volume = {23},
	year = {2023}}

@article{Dawson:2018,
	author = {Dawson, Todd E. and Goldsmith, Gregory R.},
	doi = {https://doi.org/10.1111/nph.15307},
	eprint = {https://nph.onlinelibrary.wiley.com/doi/pdf/10.1111/nph.15307},
	journal = {New Phytol.},
	number = {4},
	pages = {1156-1169},
	title = {The value of wet leaves},
	url = {https://nph.onlinelibrary.wiley.com/doi/abs/10.1111/nph.15307},
	volume = {219},
	year = {2018}}

@article{Zhou:2020,
	author = {Zhou, Xingyi and Lu, Hengyi and Zhao, Fei and Yu, Guihua},
	date = {2020/07/06},
	doi = {10.1021/acsmaterialslett.0c00130},
	journal = {ACS Materials Letters},
	journal1 = {ACS Materials Letters},
	journal2 = {ACS Materials Lett.},
	month = {07},
	number = {7},
	pages = {671--684},
	publisher = {American Chemical Society},
	title = {Atmospheric Water Harvesting: A Review of Material and Structural Designs},
	type = {doi: 10.1021/acsmaterialslett.0c00130},
	url = {https://doi.org/10.1021/acsmaterialslett.0c00130},
	volume = {2},
	year = {2020},
	year1 = {2020}}

@article{Ju:2012,
	author = {Ju, Jie and Bai, Hao and Zheng, Yongmei and Zhao, Tianyi and Fang, Ruochen and Jiang, Lei},
	date = {2012/12/04},
	doi = {10.1038/ncomms2253},
	id = {Ju2012},
	isbn = {2041-1723},
	journal = {Nature Communications},
	number = {1},
	pages = {1247},
	title = {A multi-structural and multi-functional integrated fog collection system in cactus},
	url = {https://doi.org/10.1038/ncomms2253},
	volume = {3},
	year = {2012}}

@article{Malik:2014,
	address = {The Institute of Life Science, College of Medicine, Swansea University, Swansea, SA2 8PP, UK.},
	author = {Malik, F T and Clement, R M and Gethin, D T and Krawszik, W and Parker, A R},
	crdt = {2014/03/21 06:00},
	date = {2014 Sep},
	dcom = {20150514},
	dep = {20140320},
	doi = {10.1088/1748-3182/9/3/031002},
	edat = {2014/03/22 06:00},
	issn = {1748-3190 (Electronic); 1748-3182 (Linking)},
	jid = {101292902},
	journal = {Bioinspir Biomim},
	jt = {Bioinspiration \& biomimetics},
	language = {eng},
	lid = {10.1088/1748-3182/9/3/031002 {$[$}doi{$]$}},
	lr = {20140828},
	mh = {Animals; Diffusion; Humans; Models, Biological; Models, Chemical; Plants/*chemistry/*metabolism; Skin/*chemistry/*metabolism; Skin Absorption/*physiology; Surface Properties; Water/*chemistry/*metabolism},
	mhda = {2015/05/15 06:00},
	number = {3},
	own = {NLM},
	pages = {031002},
	phst = {2014/03/21 06:00 {$[$}entrez{$]$}; 2014/03/22 06:00 {$[$}pubmed{$]$}; 2015/05/15 06:00 {$[$}medline{$]$}},
	pl = {England},
	pst = {ppublish},
	title = {Nature's moisture harvesters: a comparative review.},
	volume = {9},
	year = {2014}}

@article{Bin:2022,
	author = {Bin Rahman, A N M Rubaiyath and Ding, Wona and Zhang, Jianhua},
	doi = {10.1093/plphys/kiac179},
	eprint = {https://academic.oup.com/plphys/article-pdf/189/3/1435/44267523/kiac179.pdf},
	issn = {0032-0889},
	journal = {Plant Physiology},
	month = {04},
	number = {3},
	pages = {1435-1449},
	title = {The absorption of water from humid air by grass embryos during germination},
	url = {https://doi.org/10.1093/plphys/kiac179},
	volume = {189},
	year = {2022}}

@article{Waseem:2021,
	author = {Waseem, Muhammad and Nie, Zheng-Fei and Yao, Guang-Qian and Hasan, Mahadi and Xiang, Yun and Fang, Xiang-Wen},
	doi = {https://doi.org/10.1111/ppl.13334},
	eprint = {https://onlinelibrary.wiley.com/doi/pdf/10.1111/ppl.13334},
	journal = {Physiologia Plantarum},
	number = {2},
	pages = {528-539},
	title = {Dew absorption by leaf trichomes in : An alternative water acquisition strategy for withstanding drought in arid environments},
	url = {https://onlinelibrary.wiley.com/doi/abs/10.1111/ppl.13334},
	volume = {172},
	year = {2021}}

@article{Liu:2017,
  author={Liu, J. and Yang, H. and Gosling, S. N. and Kummu, M. and Flörke, M. and Pfister, S. and Hanasaki, N. and Wada, Y. and Zhang, X. and Zheng, C. and Alcamo, J. and Oki, T.},
  title={Water scarcity assessments in the past, present and future},
  year={2017},
  journal={Earth's Future},
  volume={5},
  number={6},
  pages={545--559},
  doi={10.1002/2016EF000518},
}

@article{Gerasopoulos:2018,
	author = {Gerasopoulos, Konstantinos and Luedeman, William L. and {\"O}l{\c c}eroglu, Emre and McCarthy, Matthew and Benkoski, Jason J.},
	date = {2018/01/31},
	doi = {10.1021/acsami.7b16379},
	isbn = {1944-8244},
	journal = {ACS Applied Materials \& Interfaces},
	journal1 = {ACS Applied Materials \& Interfaces},
	journal2 = {ACS Appl. Mater. Interfaces},
	month = {01},
	number = {4},
	pages = {4066--4076},
	publisher = {American Chemical Society},
	title = {Effects of Engineered Wettability on the Efficiency of Dew Collection},
	type = {doi: 10.1021/acsami.7b16379},
	url = {https://doi.org/10.1021/acsami.7b16379},
	volume = {10},
	year = {2018},
	year1 = {2018}}

@article{Cavallaro:2022,
	author = {Cavallaro, Agust{\'\i}n and Carbonell-Silletta, Luisina and Burek, Antonella and Goldstein, Guillermo and Scholz, Fabi{\'a}n G and Bucci, Sandra J},
	crdt = {2022/03/24 17:25},
	date = {2022 Sep 19},
	dcom = {20220922},
	doi = {10.1093/aob/mcac042},
	edat = {2022/03/25 06:00},
	issn = {1095-8290 (Electronic); 0305-7364 (Print); 0305-7364 (Linking)},
	jid = {0372347},
	journal = {Ann Bot},
	jt = {Annals of botany},
	language = {eng},
	lid = {10.1093/aob/mcac042 {$[$}doi{$]$}},
	lr = {20240904},
	mh = {Ecosystem; *Groundwater; Plant Leaves; Plants; Soil; *Water; Wettability},
	mhda = {2022/09/23 06:00},
	month = {Sep},
	number = {3},
	oto = {NOTNLM},
	own = {NLM},
	pages = {409--418},
	pii = {6553349; mcac042},
	pl = {England},
	pmc = {PMC9486909},
	pmcr = {2023/03/24},
	pmid = {35325023},
	pst = {ppublish},
	pt = {Journal Article; Research Support, Non-U.S. Gov't},
	rn = {0 (Soil); 059QF0KO0R (Water)},
	sb = {IM},
	status = {MEDLINE},
	title = {Leaf surface traits contributing to wettability, water interception and uptake of above-ground water sources in shrubs of Patagonian arid ecosystems.},
	volume = {130},
	year = {2022}}

@article{Liu:2020a,
	author = {Meizhen Liu and Yu Cen and Chengdong Wang and Xian Gu and Peter Bowler and Dongxiu Wu and Lin Zhang and Gaoming Jiang and Daniel Beysens},
	doi = {https://doi.org/10.1016/j.agrformet.2020.107941},
	issn = {0168-1923},
	journal = {Agricultural and Forest Meteorology},
	pages = {107941},
	title = {Foliar uptake of dew in the sandy ecosystem of the Mongolia Plateau: A life-sustaining and carbon accumulation strategy shared differently by C3 and C4 grasses},
	url = {https://www.sciencedirect.com/science/article/pii/S0168192320300435},
	volume = {287},
	year = {2020}}

@article{Xu:2022,
	author = {Xu, Yingying and Yang, Xu and Dou, Yingbo and Yi, Yan},
	doi = {10.15244/pjoes/148064},
	issn = {1230-1485},
	journal = {Polish Journal of Environmental Studies},
	number = {5},
	pages = {4427--4434},
	title = {The Influencing Factors of Dew Absorbed by Leaves},
	url = {https://doi.org/10.15244/pjoes/148064},
	volume = {31},
	year = {2022}}

@article{Bayer:2020,
	author = {Bayer, Ilker S.},
	doi = {https://doi.org/10.1002/admi.202000095},
	eprint = {https://onlinelibrary.wiley.com/doi/pdf/10.1002/admi.202000095},
	journal = {Adv. Mater. Interf.},
	number = {13},
	pages = {2000095},
	title = {Superhydrophobic Coatings from Ecofriendly Materials and Processes: A Review},
	url = {https://onlinelibrary.wiley.com/doi/abs/10.1002/admi.202000095},
	volume = {7},
	year = {2020}}

@article{Ensikat:2011,
	author = {Ensikat, Hans J and Ditsche-Kuru, Petra and Neinhuis, Christoph and Barthlott, Wilhelm},
	crdt = {2011/10/07 06:00},
	date = {2011},
	dcom = {20111110},
	dep = {20110310},
	doi = {10.3762/bjnano.2.19},
	edat = {2011/10/07 06:00},
	issn = {2190-4286 (Electronic); 2190-4286 (Linking)},
	jid = {101551563},
	journal = {Beilstein J. Nanotechnol.},
	jt = {Beilstein journal of nanotechnology},
	language = {eng},
	lid = {10.3762/bjnano.2.19 {$[$}doi{$]$}},
	lr = {20220318},
	mhda = {2011/10/07 06:01},
	oto = {NOTNLM},
	own = {NLM},
	pages = {152--161},
	phst = {2011/01/07 00:00 {$[$}received{$]$}; 2011/02/17 00:00 {$[$}accepted{$]$}; 2011/10/07 06:00 {$[$}entrez{$]$}; 2011/10/07 06:00 {$[$}pubmed{$]$}; 2011/10/07 06:01 {$[$}medline{$]$}},
	pmc = {PMC3148040},
	pmid = {21977427},
	pst = {ppublish},
	pt = {Journal Article},
	status = {PubMed-not-MEDLINE},
	title = {Superhydrophobicity in perfection: the outstanding properties of the lotus leaf.},
	volume = {2},
	year = {2011}}

@article{Holloway:1969,
	author = {Holloway, P. J.},
	doi = {https://doi.org/10.1111/j.1744-7348.1969.tb05475.x},
	eprint = {https://onlinelibrary.wiley.com/doi/pdf/10.1111/j.1744-7348.1969.tb05475.x},
	journal = {Ann. Appl. Biol.},
	number = {1},
	pages = {145-153},
	title = {The effects of superficial wax on leaf wettability},
	url = {https://onlinelibrary.wiley.com/doi/abs/10.1111/j.1744-7348.1969.tb05475.x},
	volume = {63},
	year = {1969}}

@article{Szczepanski:2017,
	author = {Caroline R. Szczepanski and Fr{\'e}d{\'e}ric Guittard and Thierry Darmanin},
	doi = {https://doi.org/10.1016/j.cis.2017.01.002},
	issn = {0001-8686},
	journal = {Adv. Colloid Interf. Sci.},
	pages = {37-61},
	title = {Recent advances in the study and design of parahydrophobic surfaces: From natural examples to synthetic approaches},
	url = {https://www.sciencedirect.com/science/article/pii/S0001868616302780},
	volume = {241},
	year = {2017}}

@article{Bhushan:2010,
	author = {Bharat Bhushan and Michael Nosonovsky},
	issn = {1364503X},
	journal = {Philos. T. R. Soc. A},
	number = {1929},
	pages = {4713--4728},
	publisher = {The Royal Society},
	title = {The rose petal effect and the modes of superhydrophobicity},
	url = {http://www.jstor.org/stable/25753437},
	urldate = {2023-04-17},
	volume = {368},
	year = {2010}}

@article{Chakraborty:2019,
	author = {Chakraborty, Monojit and Weibel, Justin A. and Schaber, James A. and Garimella, Suresh V.},
	doi = {https://doi.org/10.1002/admi.201900652},
	eprint = {https://onlinelibrary.wiley.com/doi/pdf/10.1002/admi.201900652},
	journal = {Adv. Mater. Interf.},
	number = {17},
	pages = {1900652},
	title = {The Wetting State of Water on a Rose Petal},
	url = {https://onlinelibrary.wiley.com/doi/abs/10.1002/admi.201900652},
	volume = {6},
	year = {2019}}

@article{Ritter:2019,
	author = {Ritter, F. and Berkelhammer, M. and Beysens, D.},
	doi = {10.5194/hess-23-1179-2019},
	journal = {Hydrology and Earth System Sciences},
	number = {2},
	pages = {1179--1197},
	title = {Dew frequency across the US from a network of in situ radiometers},
	url = {https://hess.copernicus.org/articles/23/1179/2019/},
	volume = {23},
	year = {2019}}

@article{Shepherd:2006,
	author = {Shepherd, Tom and Wynne Griffiths, D.},
	doi = {https://doi.org/10.1111/j.1469-8137.2006.01826.x},
	eprint = {https://nph.onlinelibrary.wiley.com/doi/pdf/10.1111/j.1469-8137.2006.01826.x},
	journal = {New Phytologist},
	number = {3},
	pages = {469-499},
	title = {The effects of stress on plant cuticular waxes},
	url = {https://nph.onlinelibrary.wiley.com/doi/abs/10.1111/j.1469-8137.2006.01826.x},
	volume = {171},
	year = {2006}}

@article{Tomaszkiewicz:2015,
	author = {Tomaszkiewicz, Marlene and Abou Najm, Majdi and Beysens, Daniel and Alameddine, Ibrahim and El-Fadel, Mutasem},
	doi = {10.1139/er-2015-0035},
	eprint = {https://doi.org/10.1139/er-2015-0035},
	journal = {Environmental Reviews},
	number = {4},
	pages = {425-442},
	title = {Dew as a sustainable non-conventional water resource: a critical review},
	url = {https://doi.org/10.1139/er-2015-0035},
	volume = {23},
	year = {2015}}

@article{Monteith:1965,
	author = {Monteith, J. L. and Szeicz, G. and Waggoner, P. E.},
	c1 = {Full publication date: Nov., 1965},
	db = {JSTOR},
	doi = {10.2307/2401484},
	isbn = {00218901, 13652664},
	journal = {Journal of Applied Ecology},
	month = {2024/10/30/},
	number = {2},
	pages = {345--355},
	publisher = {{$[$}British Ecological Society, Wiley{$]$}},
	title = {The Measurement and Control of Stomatal Resistance in the Field},
	url = {http://www.jstor.org/stable/2401484},
	volume = {2},
	year = {1965}}

@article{Zheng:2021,
	author = {Shao-Fei Zheng and Ulrich Gross and Xiao-Dong Wang},
	doi = {https://doi.org/10.1016/j.cis.2021.102503},
	issn = {0001-8686},
	journal = {Advances in Colloid and Interface Science},
	pages = {102503},
	title = {Dropwise condensation: From fundamentals of wetting, nucleation, and droplet mobility to performance improvement by advanced functional surfaces},
	url = {https://www.sciencedirect.com/science/article/pii/S0001868621001445},
	volume = {295},
	year = {2021}}

@article{Long:1955,
	author = {Long, I. F.},
	doi = {https://doi.org/10.1002/j.1477-8696.1955.tb00170.x},
	eprint = {https://rmets.onlinelibrary.wiley.com/doi/pdf/10.1002/j.1477-8696.1955.tb00170.x},
	journal = {Weather},
	number = {4},
	pages = {128-128},
	title = {Dew and guttation},
	url = {https://rmets.onlinelibrary.wiley.com/doi/abs/10.1002/j.1477-8696.1955.tb00170.x},
	volume = {10},
	year = {1955}}

@article{Hughes:1994,
	author = {R.N. Hughes and P. Brimblecombe},
	doi = {https://doi.org/10.1016/0168-1923(94)90002-7},
	issn = {0168-1923},
	journal = {Agricultural and Forest Meteorology},
	number = {3},
	pages = {173-190},
	title = {Dew and guttation: formation and environmental significance},
	url = {https://www.sciencedirect.com/science/article/pii/0168192394900027},
	volume = {67},
	year = {1994}}

@article{Gerlein:2018,
	author = {Cynthia Gerlein-Safdi and Michael C. Koohafkan and Michaella Chung and Fulton E. Rockwell and Sally Thompson and Kelly K. Caylor},
	doi = {https://doi.org/10.1016/j.agrformet.2018.05.015},
	issn = {0168-1923},
	journal = {Agricultural and Forest Meteorology},
	pages = {305-316},
	title = {Dew deposition suppresses transpiration and carbon uptake in leaves},
	url = {https://www.sciencedirect.com/science/article/pii/S0168192318301679},
	volume = {259},
	year = {2018}}

@article{Jacobs:2008,
	author = {A.F.G. Jacobs and B.G. Heusinkveld and S.M. Berkowicz},
	doi = {https://doi.org/10.1016/j.atmosres.2007.06.007},
	issn = {0169-8095},
	journal = {Atmospheric Research},
	note = {Third International Conference on Fog, Fog Collection and Dew},
	number = {3},
	pages = {377-385},
	title = {Passive dew collection in a grassland area, The Netherlands},
	url = {https://www.sciencedirect.com/science/article/pii/S0169809507001950},
	volume = {87},
	year = {2008}}

@article{Nath:2019,
	author = {Nath, Saurabh and Ahmadi, S. Farzad and Gruszewski, Hope A. and Budhiraja, Stuti and Bisbano, Caitlin E. and Jung, Sunghwan and Schmale, David G. and Boreyko, Jonathan B.},
	doi = {10.1098/rsif.2019.0243},
	eprint = {https://royalsocietypublishing.org/doi/pdf/10.1098/rsif.2019.0243},
	journal = {Journal of The Royal Society Interface},
	number = {155},
	pages = {20190243},
	title = {`Sneezing' plants: pathogen transport via jumping-droplet condensation},
	url = {https://royalsocietypublishing.org/doi/abs/10.1098/rsif.2019.0243},
	volume = {16},
	year = {2019}}

@article{Li:2002,
	author = {Xiao-Yan Li},
	doi = {https://doi.org/10.1016/S0022-1694(01)00605-9},
	issn = {0022-1694},
	journal = {Journal of Hydrology},
	number = {1},
	pages = {151-160},
	title = {Effects of gravel and sand mulches on dew deposition in the semiarid region of China},
	url = {https://www.sciencedirect.com/science/article/pii/S0022169401006059},
	volume = {260},
	year = {2002}}

@article{Zhuang:2017,
	author = {Yanli Zhuang and Wenzhi Zhao},
	doi = {https://doi.org/10.1016/j.agrformet.2017.08.032},
	issn = {0168-1923},
	journal = {Agricultural and Forest Meteorology},
	pages = {541-550},
	title = {Dew formation and its variation in Haloxylon ammodendron plantations at the edge of a desert oasis, northwestern China},
	url = {https://www.sciencedirect.com/science/article/pii/S0168192317302897},
	volume = {247},
	year = {2017}}

@article{Prochazka:2024,
	author = {Prochazka, Luana S. and Alcantara, Suzana and Rando, Juliana Gastaldello and Vasconcelos, Thais and Pizzardo, Raquel C. and Nogueira, Anselmo},
	doi = {https://doi.org/10.1111/nph.19601},
	eprint = {https://nph.onlinelibrary.wiley.com/doi/pdf/10.1111/nph.19601},
	journal = {New Phytologist},
	number = {2},
	pages = {760-773},
	title = {Resource availability and disturbance frequency shape evolution of plant life forms in Neotropical habitats},
	url = {https://nph.onlinelibrary.wiley.com/doi/abs/10.1111/nph.19601},
	volume = {242},
	year = {2024}}

@article{Bobe:2006,
	author = {R. Bobe},
	doi = {https://doi.org/10.1016/j.jaridenv.2006.01.010},
	issn = {0140-1963},
	journal = {Journal of Arid Environments},
	note = {Special Issue Historical biogeography and origin and evolution of arid and semi-arid environments},
	number = {3},
	pages = {564-584},
	title = {The evolution of arid ecosystems in eastern Africa},
	url = {https://www.sciencedirect.com/science/article/pii/S0140196306000425},
	volume = {66},
	year = {2006}}

@article{Rockwell:2022,
	address = {Department of Organismal Biology, Harvard University, Cambridge, MA, USA.; Department of Ecology and Evolutionary Biology, The University of Toronto, 25 Willcocks Street, Toronto, ON M5S3B2, Canada.},
	author = {Rockwell, Fulton and Sage, Rowan F},
	copyright = {{\copyright}The Author(s) 2022. Published by Oxford University Press on behalf of the Annals of Botany Company. All rights reserved. For permissions, please e-mail: journals.permissions@oup.com.},
	crdt = {2022/08/23 11:22},
	date = {2022 Sep 19},
	dcom = {20220922},
	doi = {10.1093/aob/mcac107},
	edat = {2022/08/24 06:00},
	issn = {1095-8290 (Electronic); 0305-7364 (Print); 0305-7364 (Linking)},
	jid = {0372347},
	journal = {Ann Bot},
	jt = {Annals of botany},
	language = {eng},
	lid = {10.1093/aob/mcac107 {$[$}doi{$]$}},
	lr = {20240904},
	mh = {Carbon; *Plants; *Water},
	mhda = {2022/09/23 06:00},
	month = {Sep},
	number = {3},
	own = {NLM},
	pages = {i-viii},
	phst = {2022/08/11 00:00 {$[$}received{$]$}; 2022/08/14 00:00 {$[$}accepted{$]$}; 2022/08/24 06:00 {$[$}pubmed{$]$}; 2022/09/23 06:00 {$[$}medline{$]$}; 2022/08/23 11:22 {$[$}entrez{$]$}; 2023/08/23 00:00 {$[$}pmc-release{$]$}},
	pii = {6673937; mcac107},
	pl = {England},
	pmc = {PMC9486925},
	pmcr = {2023/08/23},
	pmid = {35997781},
	pst = {ppublish},
	pt = {Journal Article},
	rn = {059QF0KO0R (Water); 7440-44-0 (Carbon)},
	sb = {IM},
	status = {MEDLINE},
	title = {Plants and water: the search for a comprehensive understanding.},
	volume = {130},
	year = {2022}}

@article{Lusa:2014,
	author = {Lusa, Makeli Garibotti and Cardoso, Elaine Cristina and Machado, Silvia Rodrigues and Appezzato-da-Gl{\'o}ria, Beatriz},
	crdt = {2014/12/21 06:00},
	date = {2014 Dec 19},
	dcom = {20150128},
	dep = {20141219},
	doi = {10.1093/aobpla/plu088},
	edat = {2014/12/21 06:00},
	issn = {2041-2851 (Print); 2041-2851 (Electronic)},
	jid = {101539425},
	journal = {AoB Plants},
	jt = {AoB PLANTS},
	language = {eng},
	lid = {plu088 {$[$}pii{$]$}; 10.1093/aobpla/plu088 {$[$}doi{$]$}},
	lr = {20200930},
	mhda = {2014/12/21 06:01},
	month = {Dec},
	oto = {NOTNLM},
	own = {NLM},
	phst = {2014/12/21 06:00 {$[$}entrez{$]$}; 2014/12/21 06:00 {$[$}pubmed{$]$}; 2014/12/21 06:01 {$[$}medline{$]$}; 2015/01/01 00:00 {$[$}pmc-release{$]$}},
	pii = {plu088},
	pl = {England},
	pmc = {PMC4381741},
	pmcr = {2015/01/01},
	pmid = {25527474},
	pst = {epublish},
	pt = {Journal Article},
	status = {PubMed-not-MEDLINE},
	title = {Trichomes related to an unusual method of water retention and protection of the stem apex in an arid zone perennial species.},
	volume = {7},
	year = {2014}}

@article{Barthlott:2016,
	author = {Barthlott, W. and Mail, M. and Neinhuis, C.},
	doi = {10.1098/rsta.2016.0191},
	eprint = {https://royalsocietypublishing.org/doi/pdf/10.1098/rsta.2016.0191},
	journal = {Philos. T. R. Soc. A},
	number = {2073},
	pages = {20160191},
	title = {Superhydrophobic hierarchically structured surfaces in biology: evolution, structural principles and biomimetic applications},
	url = {https://royalsocietypublishing.org/doi/abs/10.1098/rsta.2016.0191},
	volume = {374},
	year = {2016}}

@article{Andrews:2011,
	author = {Andrews, H. G. and Eccles, E. A. and Schofield, W. C. E. and Badyal, J. P. S.},
	date = {2011/04/05},
	doi = {10.1021/la2000014},
	isbn = {0743-7463},
	journal = {Langmuir},
	journal1 = {Langmuir},
	journal2 = {Langmuir},
	month = {04},
	number = {7},
	pages = {3798--3802},
	publisher = {American Chemical Society},
	title = {Three-Dimensional Hierarchical Structures for Fog Harvesting},
	type = {doi: 10.1021/la2000014},
	url = {https://doi.org/10.1021/la2000014},
	volume = {27},
	year = {2011},
	year1 = {2011}}

@article{Mauseth:2006,
	address = {Section of Integrative Biology, 1 University Station, A6700, University of Texas, Austin, TX 78712, USA. j.mauseth@mail.utexas.edu},
	author = {Mauseth, James D},
	crdt = {2006/07/06 09:00},
	date = {2006 Nov},
	dcom = {20061219},
	dep = {20060704},
	doi = {10.1093/aob/mcl133},
	edat = {2006/07/06 09:00},
	issn = {0305-7364 (Print); 1095-8290 (Electronic); 0305-7364 (Linking)},
	jid = {0372347},
	journal = {Ann Bot},
	jt = {Annals of botany},
	language = {eng},
	lr = {20181220},
	mh = {*Cactaceae; *Plant Shoots; Structure-Activity Relationship},
	mhda = {2006/12/21 09:00},
	month = {Nov},
	number = {5},
	own = {NLM},
	pages = {901--926},
	phst = {2006/07/06 09:00 {$[$}pubmed{$]$}; 2006/12/21 09:00 {$[$}medline{$]$}; 2006/07/06 09:00 {$[$}entrez{$]$}; 2007/11/01 00:00 {$[$}pmc-release{$]$}},
	pii = {mcl133},
	pl = {England},
	pmc = {PMC2803597},
	pmcr = {2007/11/01},
	pmid = {16820405},
	pst = {ppublish},
	pt = {Journal Article; Research Support, Non-U.S. Gov't; Review},
	sb = {IM},
	status = {MEDLINE},
	title = {Structure-function relationships in highly modified shoots of cactaceae.},
	volume = {98},
	year = {2006}}

@article{Gao:2024,
	author = {Yiwei Gao and Areianna Eason and Santiago Ricoy and Addison Cobb and Ryan Phung and Amir Kashani and Mario R. Mata and Aaron Sahm and Nathan Ortiz and Sameer Rao and H. Jeremy Cho},
	doi = {10.1073/pnas.2321429121},
	eprint = {https://www.pnas.org/doi/pdf/10.1073/pnas.2321429121},
	journal = {Proceedings of the National Academy of Sciences},
	number = {44},
	pages = {e2321429121},
	title = {High-yield atmospheric water capture via bioinspired material segregation},
	url = {https://www.pnas.org/doi/abs/10.1073/pnas.2321429121},
	volume = {121},
	year = {2024}}

@article{Bhushan:2019,
	author = {Bhushan, Bharat},
	doi = {10.1098/rsta.2019.0119},
	eprint = {https://royalsocietypublishing.org/doi/pdf/10.1098/rsta.2019.0119},
	journal = {Philosophical Transactions of the Royal Society A: Mathematical, Physical and Engineering Sciences},
	number = {2150},
	pages = {20190119},
	title = {Bioinspired water collection methods to supplement water supply},
	url = {https://royalsocietypublishing.org/doi/abs/10.1098/rsta.2019.0119},
	volume = {377},
	year = {2019}}

@article{Dai:2018,
	author = {Xianming Dai and Nan Sun and Steven O. Nielsen and Birgitt Boschitsch Stogin and Jing Wang and Shikuan Yang and Tak-Sing Wong},
	doi = {10.1126/sciadv.aaq0919},
	eprint = {https://www.science.org/doi/pdf/10.1126/sciadv.aaq0919},
	journal = {Science Advances},
	number = {3},
	pages = {eaaq0919},
	title = {Hydrophilic directional slippery rough surfaces for water harvesting},
	url = {https://www.science.org/doi/abs/10.1126/sciadv.aaq0919},
	volume = {4},
	year = {2018}}

@article{Sharma:1976,
	author = {Sharma, M.L},
	doi = {10.1016/0002-1571(76)90086-8},
	journal = {Agricultural Meteorology},
	note = {Cited by: 36},
	number = {5},
	pages = {321 -- 331},
	publication_stage = {Final},
	source = {Scopus},
	title = {Contribution of dew in the hydrologic balance of a semi-arid grassland},
	type = {Article},
	volume = {17},
	year = {1976}}

@article{Agam:2006,
	author = {N. Agam and P.R. Berliner},
	doi = {https://doi.org/10.1016/j.jaridenv.2005.09.004},
	issn = {0140-1963},
	journal = {Journal of Arid Environments},
	number = {4},
	pages = {572-590},
	title = {Dew formation and water vapor adsorption in semi-arid environments---A review},
	url = {https://www.sciencedirect.com/science/article/pii/S0140196305002235},
	volume = {65},
	year = {2006}}

@article{Muselli:2009,
	author = {M. Muselli and D. Beysens and M. Mileta and I. Milimouk},
	doi = {https://doi.org/10.1016/j.atmosres.2009.01.004},
	issn = {0169-8095},
	journal = {Atmospheric Research},
	number = {4},
	pages = {455-463},
	title = {Dew and rain water collection in the Dalmatian Coast, Croatia},
	url = {https://www.sciencedirect.com/science/article/pii/S0169809509000209},
	volume = {92},
	year = {2009}}

@article{Hirst:1954,
	author = {Hirst, J. M.},
	doi = {https://doi.org/10.1002/qj.49708034410},
	eprint = {https://rmets.onlinelibrary.wiley.com/doi/pdf/10.1002/qj.49708034410},
	journal = {Quarterly Journal of the Royal Meteorological Society},
	number = {344},
	pages = {227-231},
	title = {A method for recording the formation and persistence of water deposits on plant shoots},
	url = {https://rmets.onlinelibrary.wiley.com/doi/abs/10.1002/qj.49708034410},
	volume = {80},
	year = {1954}}

@article{Eglinton:1967,
	author = {Geoffrey Eglinton and Richard J. Hamilton},
	eprint = {https://www.science.org/doi/pdf/10.1126/science.156.3780.1322},
	journal = {Science},
	number = {3780},
	pages = {1322-1335},
	title = {Leaf Epicuticular Waxes},
	volume = {156},
	year = {1967}}

@article{Koch:2008,
	author = {Koch, Kerstin and Ensikat, Hans-J{\"u}rgen},
	date = {2008/10/01/},
	doi = {https://doi.org/10.1016/j.micron.2007.11.010},
	isbn = {0968-4328},
	journal = {Micron.},
	number = {7},
	pages = {759--772},
	title = {The hydrophobic coatings of plant surfaces: Epicuticular wax crystals and their morphologies, crystallinity and molecular self-assembly},
	url = {https://www.sciencedirect.com/science/article/pii/S0968432807002016},
	volume = {39},
	year = {2008}}

@article{Joanny:1990,
	author = {Joanny, J. F. and Robbins, Mark O.},
	doi = {10.1063/1.458579},
	eprint = {https://pubs.aip.org/aip/jcp/article-pdf/92/5/3206/18984770/3206\_1\_online.pdf},
	issn = {0021-9606},
	journal = {The Journal of Chemical Physics},
	month = {03},
	number = {5},
	pages = {3206-3212},
	title = {Motion of a contact line on a heterogeneous surface},
	url = {https://doi.org/10.1063/1.458579},
	volume = {92},
	year = {1990}}

@article{Miljkovic:2013,
	author = {Miljkovic, Nenad and Enright, Ryan and Nam, Youngsuk and Lopez, Ken and Dou, Nicholas and Sack, Jean and Wang, Evelyn N.},
	date = {2013/01/09},
	doi = {10.1021/nl303835d},
	isbn = {1530-6984},
	journal = {Nano Letters},
	journal1 = {Nano Letters},
	journal2 = {Nano Lett.},
	month = {01},
	number = {1},
	pages = {179--187},
	publisher = {American Chemical Society},
	title = {Jumping-Droplet-Enhanced Condensation on Scalable Superhydrophobic Nanostructured Surfaces},
	type = {doi: 10.1021/nl303835d},
	url = {https://doi.org/10.1021/nl303835d},
	volume = {13},
	year = {2013},
	year1 = {2013}}

@article{Jonathan:2024,
	author = {Boreyko, Jonathan B.},
	doi = {https://doi.org/10.1002/dro2.105},
	eprint = {https://onlinelibrary.wiley.com/doi/pdf/10.1002/dro2.105},
	journal = {Droplet},
	number = {2},
	pages = {e105},
	title = {Jumping droplets},
	url = {https://onlinelibrary.wiley.com/doi/abs/10.1002/dro2.105},
	volume = {3},
	year = {2024}}

@article{Broza:1979,
	author = {Meir Broza},
	doi = {https://doi.org/10.1016/S0140-1963(18)31703-8},
	issn = {0140-1963},
	journal = {Journal of Arid Environments},
	number = {1},
	pages = {43-49},
	title = {Dew, fog and hygroscopic food as a source of water for desert arthropods},
	url = {https://www.sciencedirect.com/science/article/pii/S0140196318317038},
	volume = {2},
	year = {1979}}

@article{Zhu:2013,
	author = {Xiaoyi Zhu and Lizhong Xiong},
	doi = {10.1073/pnas.1316412110},
	eprint = {https://www.pnas.org/doi/pdf/10.1073/pnas.1316412110},
	journal = {Proceedings of the National Academy of Sciences},
	number = {44},
	pages = {17790-17795},
	title = {Putative megaenzyme DWA1 plays essential roles in drought resistance by regulating stress-induced wax deposition in rice},
	url = {https://www.pnas.org/doi/abs/10.1073/pnas.1316412110},
	volume = {110},
	year = {2013}}
\end{document}